\documentclass[sort&compress,preprint,review,12pt]{elsarticle}

\usepackage{amssymb}
\usepackage{amsthm}
\usepackage{caption}
\usepackage{subcaption}

\newtheorem{thm}{Theorem}

\newdefinition{example}{Example}
\newdefinition{rmk}{Remark}
\newproof{pf}{Proof}

\usepackage{amsmath}
\usepackage[makeroom]{cancel}

\usepackage{color}
\usepackage{hyperref}

\usepackage{ulem} 

\usepackage[T1]{fontenc}        

\def\QED{\mbox{\rule[0pt]{1.0ex}{1.0ex}}} 
\def\proof{\noindent\hspace{2em}{\it Proof: }}
\def\endproof{\hspace*{\fill}~\QED\par\endtrivlist\unskip}

\def\re{{\rm I}\! {\rm R}}

\begin{document}

\begin{frontmatter}



\title{Dynamic Smooth Sliding Control Applied to UAV Trajectory Tracking}


\author[PEE]{Alessandro Jacoud Peixoto}
\ead{jacoud@poli.ufrj.br}

\author[PEE]{Wenderson G. Serrantola}

\author[PEE]{Fernando~Lizarralde}

\affiliation[PEE]{organization={Department of Electrical Engineering, Federal University of Rio de Janeiro (COPPE/UFRJ)},
            city={Rio de Janeiro},
            country={Brazil}}

\begin{abstract}
This paper proposes a sliding mode controller with smooth control effort for a class of nonlinear plants. The proposed controller is created by allowing some constant parameters of the earlier smooth sliding control (SSC) to vary as a function of the output tracking error, improving the control chattering alleviation in practical implementations. Furthermore, during the sliding mode, the new scheme can synthesize a range of controllers, such as fixed gain PI controllers and approximations of the standard Super-Twisting Algorithm (STA), as well as, of the variable gain Super-Twisting Algorithm (VGSTA). A complete closed-loop stability analysis is provided. In addition, realistic simulation results with an unmanned aerial vehicle (UAV) model, incorporating  aerodynamic effects and  internal closed-loop controllers, are obtained and validated via experiments with a commercial hexacopter.
\end{abstract} 


\begin{highlights}
\item A dynamic smooth sliding  control (DSSC) with chattering avoidance and global stability properties is designed as a generalization that incorporates  functions depending on the tracking error in a previous smooth sliding  control (SSC) scheme. 

\item The control algorithm employs a smooth filter and an internal predictor that assure an ideal sliding mode in the internal predictor, chattering avoidance, and smooth control effort, at the same time. 

\item The DSSC's dynamic functions can be designed so that the synthesized control law during sliding mode generates a family of controllers, in particular, approximations for the standard  and  variable gain STA. 

\item The DSSC and the STA are designed here for relative degree one plants. Despite being well known that the STA suffers in the presence of unmodelled dynamics (as well as the DSSC), interestingly, the presence of the smooth filter in the DSSC allows a superior closed-loop performance  when compared with the STA.

\item Numerical simulations with a UAV dynamic model including aerodynamics effects and inner controllers and experimental evaluation with the DJIM600 commercial hexacopter illustrate the trajectory tracking performance of the proposed method.
\end{highlights}


\end{frontmatter}

\section{Introduction}
\label{sec:introduction}

Several control systems were proposed for the trajectory tracking problem of unmanned aerial vehicles (UAVs). Classic techniques such as the  Proportional-Integral-Derivative (PID) have been applied to most existing flight control systems, because of their simple design and implementation, when is reasonable to approximate the vehicle dynamics to a linear model \cite{abdelhay2019modeling,riviere2018agile,pounds2012stability}. When the linearization process \cite{bao2021design} needs to be avoided, as expected, non-linear control methods come into play, such as linear quadratic regulation control (LQR) \cite{priyambodo2020model, farrell2019error}, feedback linearization control schemes  \cite{ermeydan2021feedback, atmeh2011design} and robust strategies via sliding modes \cite{morenoSTA1}, \cite{mendoza2019continuous}. 

It is well-known that sliding mode-based controllers are robust with respect to bounded external disturbances and parameter uncertainties but suffer from the chattering phenomenon. In this context, aiming to avoid chattering, sliding mode control based on the STA has been widely applied \cite{morenoSTA2,morenoVGSTA2,morenoAdaptiveSTA, castillo2016super,moreno2012strict,morenoVGSTA1,haimovich2022generalized}. 
For the trajectory tracking problem of UAVs, in  \cite{gonzalez2022real,gonzalez2021observer}, the altitude control of a quadrotor is based on a combined STA and high-order sliding mode (HOSM) observer, and in \cite{tripathi2020finite}, a similar combination is addressed  to estimate  linear,  angular velocities, and unknown lumped disturbance and control the DJIMatrice 100. In order to be able to cover inspection tasks with UAVs, manipulators have been attached, such as in \cite{anuar2022super}, where an STA with gain adaptation law is designed independently of the disturbance bound caused by the manipulator dynamics.

Additionally, beyond affecting UAV's dynamics by adding manipulators, UAV geometric parameters can be variable over time \cite{derrouaoui2021nonlinear} and  the pick-and-place task can generate mass variation \cite{kim2020sliding}, \cite{peixoto2019smooth}. In \cite{derrouaoui2021nonlinear}, a Fast Terminal Sliding Mode Controller was applied to guarantee the flight stability and rapid convergence of the variable states in finite time with a reconfigurable UAV. In \cite{kim2020sliding}, a sliding mode technique was proposed allowing the UAV to adapt to the altered mass without re-tuning the controller and, in \cite{peixoto2019smooth}, the pick-and-place task was considered, by using the so-called Smooth Sliding Control (SSC) scheme.

The SSC was presented in \cite{hsu1997smooth}, as an alternative modification in the Variable Structure Model Reference Control (VS-MRAC) \cite{HLA:97} \cite{HC:89}, to provide a smooth control effort, since the VS-MRAC is a VSC strategy with discontinuous control effort. It introduces an averaging filter to obtain a continuous control signal. To compensate for the phase lag added by the averaging filter, an internal prediction loop was employed so that the ideal sliding mode could be preserved, leading to chattering avoidance and robustness with respect to unmodelled dynamics \cite{oliveira2022smooth}. 
Recently, this strategy was generalized for plants with time-varying control gain and applied to the autonomous landing problem in a moving platform \cite{peixoto2019smooth}. More recently, in  \cite{serrantola2022modified} was presented a new SSC scheme with the averaging filter time constant being updated via the tracking error for a real UAV trajectory tracking application.

In this paper, a novel modification of the results of \cite{serrantola2022modified} is proposed. As the main contributions of the paper, we consider:
(i) the development of the new (dynamic) SSC scheme, named DSSC, where the original fixed parameters of the SSC averaging filter and the predictor are replaced by dynamic  functions updated via the tracking error; 
(ii) a complete closed-loop stability analysis of the DSSC algorithm for the considered class of  non-linear plants;
(iii) a clear connection of the synthesized DSSC during sliding mode with the variable gain STA (VGSTA) and the standard STA, by selecting appropriate dynamic functions in the DSSC; and 
(iv) experimental evaluation of the DSSC and STA for trajectory tracking in  a real-world scenario with the DJI M600 Pro hexacopter.
In addition, we have  implemented a UAV's simulation model including the main aerodynamic effects, which was validated with the available commercial simulator. 

\subsection{Notations and Terminologies}

In general, for a scalar composite function  $f(e(t), \sigma(t),t)$, where $e(t)$ and $\sigma(t)$ are scalar functions of the time instant $t\geq 0$ (time-varying functions), we perform along the paper the abuse of notation $f(t)=f(e(t), \sigma(t),t)$.  A mixed time-domain and frequency-domain notation will be adopted in order to avoid clutter. In this manner, a rational function $G(s)$ will denote either an operator,
 where $s$ is the differential operator, or a transfer function,
 where $s$ is the Laplace complex frequency variable. Therefore, the time and frequency dependencies of the signals will be mostly omitted.

\section{Problem Formulation for Trajectory Tracking Control}\label{sec:problemfomulation}

Consider the following class of uncertain 
plants given by 
%
\begin{eqnarray}
\dot{\eta}(t) &=& A_\eta \eta(t) + B_\eta x_2(t)\,, \label{eq:planteta}\\
\dot{x}_1(t) &=& x_2(t)\,,\label{eq:plantx1} \\
\dot{x}_2(t) &=& -a_p x_2(t) - C_\eta \eta+ k_p [u_p(t)+d(y,\dot{y},t)]\,, \label{eq:plantx2}\\
y(t) &=& x_1(t)\,, \label{eq:plantoutput}
\end{eqnarray}
where $u_p \in \re$ is the control input, $y \in \re$ is the plant output, $\eta \in \re^{n-2}$ is the \textit{inverse system} (zero dynamics) state vector, $d \in \re$ is regarded as a matched input  disturbance, $k_p>0$ 
is the uncertain 
high-frequency gain (HFG), 
$a_p$ is an uncertain parameter, 
and  
$$x:=\left[\begin{array}{cc} x_1 & x_2 \end{array}\right]^T:=\left[\begin{array}{cc} y & \dot{y} \end{array}\right]^T \in \re^{2}$$
is the state vector. Without lost of generality, consider that ($A_\eta, B_\eta$) is in the canonical controllable form with $B_\eta=\left[\begin{array}{cccc} 0 & \ldots & 0&1\end{array}\right]^T \in \re^{n-2}$. We assume that $A_\eta$ is a Hurwitz matrix (minimum phase assumption) and $\eta$ {\bf is unavailable for feedback}. The uncertain function $d(y,\dot{y},t)$ is assumed piecewise continuous in $t$ and locally Lipschitz continuous in the other argument. For each solution of (\ref{eq:plantx1}) and (\ref{eq:plantx2}), there exists a maximal time interval of definition given by $[0,t_M)$, where $t_M$ may be finite or infinite. Thus, finite-time escape is not precluded \textit{a priori}. 

\begin{rmk}{\bf (Application: UAV's Trajectory Tracking)}
In Section~\ref{sec:UAV}, we describe how the class of nonlinear plants (\ref{eq:planteta})--(\ref{eq:plantoutput}) encompass the UAV application. The main point here is that we assume the existence of inner velocity controllers responsible for decoupling the four UAV degrees of freedom (linear velocities $v_x,v_y,v_z$ and yaw angle rate $\dot{\psi}$). These controllers generate the torques and forces commands via propellers' thrusts for tracking the velocity command inputs ($u_x,u_y,u_z$ or $u_\psi$).  This approach differs from many works which focus on the controller generating torques and forces to the UAV. The motivation behind this approach is the experiments conducted with the DJI M600 hexacopter, where the internal controller is not accessible.

For each degree of freedom, the resulting closed-loop system with the inner controllers has a relative degree one from the velocity command to the actual velocity when low/medium velocity profiles are considered, see Section~\ref{sec:UAV} for details. 
In (\ref{eq:planteta})--(\ref{eq:plantoutput}), $y$  is a generic output representing a UAV's degree of freedom ($p_x,p_y,p_z$ or $\psi$) and $u_p$ is the corresponding generic velocity command input  ($u_x,u_y,u_z$ or $u_\psi$).\endproof
\end{rmk}

\subsection{Control Objective}

The aim is to achieve at least semi-global convergence properties in the sense of uniform signal boundedness and asymptotic output practical tracking. The control objective is to design a control law $u_p(t)$ for the uncertain  plant (\ref{eq:plantx1})--(\ref{eq:plantoutput}) such that $y(t)$ tracks a bounded desired trajectory  $y_m(t)$ as close as possible, i.e.,  the tracking error 
\begin{eqnarray}
e(t) &:=& y(t)-y_m(t)\,,\label{eq:defe} 
\end{eqnarray}
converges to zero as $t \rightarrow +\infty$, or at least, to the neighbourhood  of zero (practical tracking). The \textit{desired trajectory} $y_m(t)$ is assumed to be smooth enough so that 
$\dot{y}_m$ and $\ddot{y}_m$ are well defined \textit{available} signals. 

\subsection{Main Assumptions}

We assume that $y$ and  $\dot{y}$ are available for feedback, so that 
\begin{equation}
\sigma_y(t) =\dot{y}(t)+l_0 y(t)\,, 
\label{eq:defsigmay}
\end{equation}
is an additional {\bf measured} output, where $l_0>0$ is a design constant. 
In this case,  the plant has relative degree one from $u$ to both $\dot{y}$ and $\sigma_y$. 
Based on measurements of $y$ and $\dot{y}$, let the \textit{relative degree one output variable} $\sigma(y,\dot{y},t):\re^3 \rightarrow \re$ be defined by
\begin{equation}
    \sigma:=\dot{e}+l_0 e = \sigma_y - \sigma_m\,, \quad \sigma_m:=\dot{y}_m+l_0 y_m\,. \label{eq:defsigma}
\end{equation}
The main idea is to design $u_p$ so that $\sigma$ tends to zero as $t \rightarrow +\infty$, or at least, to the vicinity of zero, despite the input disturbance $d(y,\dot{y},t)$. Thus, the convergence of the tracking error to a residual set is assured by setting $l_0>0$, according to (\ref{eq:defsigma}). 
%
%
%
The plant parameters $k_p$ and $a_p$ in (\ref{eq:plantx1})--(\ref{eq:plantoutput}) are assumed uncertain with known bounds and we consider a class of input disturbances that can be partitioned as
\begin{equation}
    d(y,\dot{y},t):=d_1(y,\dot{y},t)+d_2(y,t) + d_3(t)\,.\label{eq:defdpartition}
\end{equation}
The following assumption is considered:
\begin{description}
\item[{\bf (A0.a)}] There exist positive constants $\underline{k}_p$, $\bar{k}_p$ and $\bar{a}_p$, such that 
$$0<\underline{k}_p  \leq |k_p| \leq \bar{k}_p\quad \mbox{and} \quad |a_p| \leq \bar{a}_p\,,$$
where $\underline{k}_p$ and $\bar{a}_p$ are known constants. 
\item[{\bf (A0.b)}] There exists a known non-negative scalar function  $\alpha_d(y,\dot{y},t): \re^3 \rightarrow \re^+$, 
locally Lipschitz 
in $y$ and $\dot{y}$, piecewise continuous and upperbounded in $t$ such that 
the input disturbance $d(y,\dot{y},t)$  in (\ref{eq:defdpartition}) satisfies
$$|d(y,\dot{y},t)| \leq \alpha_d(y,\dot{y},t)\,, \quad \forall y, \dot{y}\,, \quad \forall t \in [0,t_M)\,,$$
with $\alpha_d(y,\dot{y},t) < \alpha_\sigma(|\sigma|) + \alpha_e(|e|)+\alpha_t(t)$, where $\alpha_\sigma, \alpha_e$ are class-$\mathcal{K}$ functions and $\alpha_t$ is a non-negative scalar function upperbounded in $t$.
\end{description}
Regarding the input disturbance partition (\ref{eq:defdpartition}), we also assume that:
\begin{description}
\item[{\bf (A1)}] There exist known constants $k_{d1}\geq 0$, $k_{d2}\geq 0$ and $k_{d3}\geq 0$ and known a non-negative scalar function $\alpha_{d1}(y,\dot{y},t): \re^3 \rightarrow \re^+$, 
locally Lipschitz 
in $y$ and $\dot{y}$, piecewise continuous and upperbounded in $t$ such that 
the term $d_1(y,\dot{y},t)$ in (\ref{eq:defdpartition}) satisfy
$$|d_1(y,\dot{y},t)| \leq \alpha_{d1}(y,\dot{y},t)|\sigma|\,, \quad \forall y, \dot{y}\,, \quad \forall t \in [0,t_M)\,,$$
with  $\alpha_{d1}(y,\dot{y},t):=k_{d1} |y|+ k_{d2} |\dot{y}|+k_{d3}$ and $\sigma(y,\dot{y},t)$ in (\ref{eq:defsigma}).
 
\item[{\bf (A2)}]  There exist known constants $k_{d4} \geq 0$ and $k_{d5}\geq 0$ and a known non-negative scalar function  $\alpha_{d2}(t): \re \rightarrow \re^+$,
piecewise continuous and upperbounded in $t$, such that the term $d_2(y,t)$ in (\ref{eq:defdpartition}) satisfies
$$\left|\frac{\partial d_2(y,t)}{\partial y}\right| \leq k_{d4}\quad \mbox{and} \quad \left|\frac{\partial d_2(y,t)}{\partial t}\right| \leq k_{d5} |y|+\alpha_{d2}(t)\,, \quad \forall y\,, \quad \forall t \in [0,t_M)\,.$$
\item[{\bf (A3)}] There exists a known non-negative scalar function  $\alpha_{d3}(t): \re \rightarrow \re^+$, piecewise continuous and upperbounded in $t$, such that the time derivative of the term $d_3(t)$ in (\ref{eq:defdpartition}) satisfies
$|\dot{d}_3(t)| \leq \alpha_{d3}(t)$, $\forall t \in [0,t_M)$.
\end{description}
%

%

\begin{rmk}{\bf (Plant Input Disturbance: UAV's Application)}
Regarding the application, in (\ref{eq:plantx1})--(\ref{eq:plantoutput}), the input disturbance $d(y,\dot{y},t)$ represents the coupling between the subsystems, the wind influence, and possible nonlinearities remaining due to some unmatched parameters in the inner feedback linearization action. Moreover, it is considered that the wind velocity has low-frequency components or can be represented by piecewise functions with jump discontinuities where the discontinuity points have zero measure. 
\endproof
\end{rmk}

\subsection{Error Dynamics}

In practical applications there exists some level of knowledge of the plant parameters and, in general, a nominal control  based on this knowledge is applied in conjunction with the robust action (here being the DSSC) designed to deal with disturbances and/or parameter uncertainties. Another motivation for using a nominal control $u^n$ is to reduce the DSSC's control action. 


Therefore, let the control signal be composed of two terms
\begin{equation}
  u_p(t)=u(t)+u^n(t)\,, \label{eq:updef}  
\end{equation}
where the control effort $u$ is the DSSC robust control effort and $u^n$ is a \textit{nominal control law}, both to be defined later on.

%
%
%
From (\ref{eq:defsigma}), the $e$-dynamics is directly obtained as 
\begin{equation}
    \dot{e}=-l_0 e+\sigma\,. \label{eq:defedynamics}
\end{equation}
Moreover, from (\ref{eq:defsigmay}) and (\ref{eq:defsigma}), one has $\dot{\sigma}=\dot{\sigma}_y-\dot{\sigma}_m=\ddot{y}+l_0 \dot{y}-\dot{\sigma}_m$. Moreover, from (\ref{eq:plantx1})--(\ref{eq:plantx2}) one can write $\ddot{y}=-a_p \dot{y}+k_p (u+u^n) +k_p (d-C_\eta \eta/k_p)$. Therefore, the $\sigma$-dynamics is given by 
\begin{equation}
\dot{\sigma}=k_p u+d_\sigma\,, \qquad d_\sigma:=k_p u^n +(l_0-a_p) \dot{y}-\dot{\sigma}_m+ k_p (d-C_\eta \eta/k_p) \,, \label{eq:sigmadynforsliding}
\end{equation}
where $d_\sigma$ is treated as a disturbance term.

\section{Dynamic Smooth Sliding Control (DSCC)}\label{sec:DSSC}

Despite that, the original SSC \cite{hsu1997smooth} can be applied for a broader class of plants with arbitrary relative degree \cite{peixoto2002further} and \cite{peixoto2001experimental},  we focus on the case where $y$ and $\dot{y}$ are available for feedback. 
In comparison to the original SSC \cite{hsu1997smooth} \cite{peixoto2002further}, which has fixed control parameters, the {\bf  Dynamic SSC} (DSSC) differs in one main aspect: the averaging filter time constant $\tau_{av}$, the predictor time constant $\tau_m$ and the predictor gain $k_o$ are allowed to vary with the time $t$ and/or the plant states $\sigma(t)$ and $e(t)$. 
With this modification, one can observe an improvement in control chattering alleviation in practical implementations where discretization, for instance, generates numerical chattering even for the relative degree one case.

This is improvement is achieved due to the presence of an averaging filter with a dynamic pass-band depending on the tracking error.  Far away (near) from the origin of the error system state space $(\sigma, e)$, where the tracking error is large (small), the modulation function is also large (small), and, at the same time, the dynamic pass-band is small (large). The final result is a smoother control action when compared with the original SSC.

The DSSC law is given by
\begin{equation}
u:=-u_0^{av}\,,  \label{eq:defDSSC}
\end{equation}
with a time-varying averaging  filter
\begin{eqnarray}
\tau_{av}(t) \ \dot{u}_0^{av} &= &-u_0^{av}+u_0\,,\label{eq:u0avdynamicsDSSC}
\end{eqnarray}
where $\tau_{av}(t)=\tau_{av}(\sigma(t),e(t),t) >0$ and %
\begin{equation}
u_0=\varrho(t) \ \mbox{sgn}(\tilde{\sigma})\,, \quad \varrho(t) > 0\,,  \label{eq:defu0DSSC}
\end{equation}
is the predictor's discontinuous injection term, with modulation function $\varrho(t)$. In the DSSC,  \textit{the sliding variable} $\tilde{\sigma}$ is defined as 
\begin{equation}
\tilde{\sigma}:=\sigma- \hat{\sigma}\,,
\label{eq:defsigmatildeDSSC}
\end{equation}
where $\hat{\sigma}$ is the output of the predictor 
\begin{equation}
\dot{\hat{\sigma}}=- \frac{1}{\tau_m(t) } \hat{\sigma} +  k_o(t) [-u_0^{av}+u_0]\,,
\label{eq:predictordynamicsDSSC}
\end{equation}
with $\tau_m(t)=\tau_m(\sigma(t), e(t), t) >0$ and $k_o(t)=k_o(\sigma(t), e(t), t)>0$. 
Note that, all the functions 
$$k_o(t)\,, \tau_m(t)\,, \tau_{av}(t)\,, \varrho(t) > 0\,,$$
%
(to be defined later on) can depend on exogenous time-varying functions, as well as, on the output tracking error $e$ (or $\sigma$), which in turn depends on the closed-loop system dynamics. Henceforth, we denote these functions by {\bf dynamic functions}. 
%
%

\begin{figure}[h!]
\begin{center}
\includegraphics[width=1\linewidth]{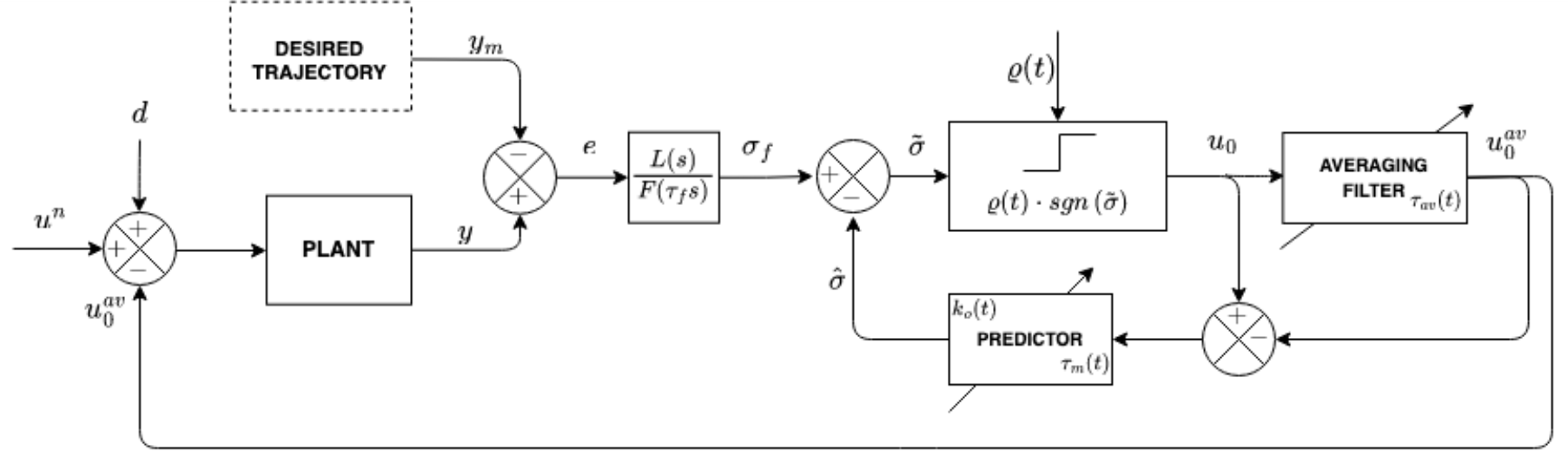}
\caption{General DSSC  block diagram for arbitrary relative degree case and with generic dynamic functions $\tau_m(\sigma(t), e(t), t)$, $k_o(\sigma(t), e(t), t)$ and $\tau_{av}(\sigma(t), e(t), t)$. The predictor is given in  (\ref{eq:predictordynamicsDSSC}) and depends on $k_o$ and $\tau_m$, while the averaging filter is given in (\ref{eq:u0avdynamicsDSSC}) and  depends on $\tau_{av}$. For the class relative degree one plants considered here with  $\dot{y}$ available for feedback,  one can set $\tau_f=0$, so that $\sigma_f=\sigma$ with $\sigma$  in (\ref{eq:defsigma}).}
\label{fig:topologiaDSSC}		
\end{center}
\end{figure}

%
%
The DSSC scheme can also deal with arbitrary relative degree plants, by using linear lead filters to estimate output time derivatives. We restrict ourselves to the case of relative degree one which is the simplest case amenable by pure Lyapunov design. More precisely, the extension to nonlinear systems with an arbitrary relative degree can be obtained by using   approximated filtered versions for $\sigma_y$, like  $\sigma_{y_f} = y_f + l_0 y$, 
where $y_f \approx \dot{y}$ is obtained by a lead filter. 
%
%
%
%
%
The  DSSC's block diagram is presented in Figure~\ref{fig:topologiaDSSC}, including the linear lead filter for completeness of the presentation. 

\subsection{Sliding Variable Dynamics: Existence of Ideal Sliding Mode}

With $\tilde{\sigma}$ defined in (\ref{eq:defsigmatildeDSSC}), the $\sigma$-dynamics in (\ref{eq:sigmadynforsliding}) and the smooth control law (\ref{eq:defDSSC}), one has that 
$\dot{\tilde{\sigma}}= \dot{\sigma}-\dot{\hat{\sigma}}=  [-k_p u_{0}^{av}+k_p u^n +(l_0-a_p) \dot{y}+ k_p (d-C_\eta \eta/k_p) -\dot{\sigma}_m]-\dot{\hat{\sigma}}$.
Moreover, by using the  predictor dynamics in (\ref{eq:predictordynamicsDSSC}) and the relationship $\hat{\sigma}=\sigma-\tilde{\sigma}$, one can further obtain 
%
\begin{eqnarray}
\tau_m \dot{\tilde{\sigma}}&=&- \tilde{\sigma}  + k_o \tau_m [-u_0+ d_0/k_o]\,,\label{eq:sigmatildedynamicsDSSC}
\end{eqnarray}
where 
\begin{eqnarray}
d_0&:=&(k_o-k_p)u_0^{av}+\tilde{d}_1+\tilde{d}_2\,,\label{eq:defd0} \\
\tilde{d}_1&:=&  
k_p u^n -a_p \dot{y}+ k_p (d-C_\eta \eta/k_p)\,, \label{eq:defdtil1}\\
\tilde{d}_2&:=& \frac{\sigma}{\tau_m} +l_0 \dot{y} -\dot{\sigma}_m\,,\label{eq:defdtil2}
\end{eqnarray}
with $\tilde{d}_1$ being an uncertain term and  $\tilde{d}_2$ being a known signal that could be directly canceled by redefining the control term  $u_0$ in (\ref{eq:defu0DSSC}). For simplicity, at the cost of being more conservatism, we treat $\tilde{d}_2$ as an uncertain term too. The investigation of canceling the term $\tilde{d}_2$ is left for future work.

As in the original SSC, sliding mode occurs at $\tilde{\sigma} \equiv 0$ so that  $\tilde{\sigma}$ converges to zero in some finite time $t_s \in  [0,t_M)$, i.e., $\hat{\sigma}(t) = \sigma(t)$, $\forall t \in [t_s,t_M)$, provided that the modulation function $\varrho$ (in the discontinuous term $u_0$) is designed properly.

The proof of  the sliding mode existence  and the avoidance of finite time escape (mainly due to the unboundedness observability property of the closed-loop system) are provided later on in Theorem~\ref{theorem:DSSC}. 
\begin{rmk}{\bf (Modulation Function Design)}\label{rem:nudef}
The modulation function is designed to dominate the norm of the total disturbance $d_0/k_o$ faced by $u_0$ in the (\ref{eq:sigmatildedynamicsDSSC}), \textit{modulo} vanishing terms due to initial conditions. The modulation function can be chosen as:
\begin{equation}
\varrho:=(k_o+\bar{k}_p) |u_0^{av}|/k_o+\tilde{D}/k_o+\delta_\rho/k_o\,,
    \label{eq:defvarrho}
\end{equation}
where 
%
\begin{equation}
    \tilde{D}:= \bar{k}_p |u^n|+ (\bar{a}_p+l_0) |\dot{y}|+ \bar{k}_p \alpha_d+|\dot{\sigma}_m|+\frac{|\sigma|}{\tau_m}+\bar{\eta}\,,\label{eq:defDTILDE}
\end{equation}
is an available norm bound for the sum $\tilde{d}_1+\tilde{d}_2$ and $\delta_\rho>0$ is an arbitrary small constant. In (\ref{eq:defDTILDE}), we have used a norm observer for the inverse system state norm to generate $\bar{\eta}>\|C_\eta \eta\|$, \textit{modulo} vanishing terms due to initial conditions, the available norm bound function $\alpha_d$ for the plant input disturbance $d$, given in {\bf (A1)},  and the available upper bounds $\bar{k}_p$ and $\bar{a}_p$ for the HFG $k_p$ and for the plant parameter $a_p$, respectively, both given in {\bf (A0)}.
Eventually, when some plant parameters are known, the magnitude of the sum $\tilde{d}_1+\tilde{d}_2$  can be reduced by choosing  $u^n$ properly or by treating $\tilde{d}_2$ as a known term, as mentioned before. \endproof
\end{rmk}

\subsection{Synthesized Equivalent Controller During Sliding Mode}

Now, let us find the synthesized equivalent DSSC control law during sliding mode. First, denote $\bar{u}_0^{av}=u_0^{av}$ as the solution of (\ref{eq:u0avdynamicsDSSC}), when  the discontinuous control $u_0$ is replaced by the  \textit{equivalent control} $u_{eq}=\frac{d_0}{k_o}$, directly obtained from the $\tilde{\sigma}$-dynamics (\ref{eq:sigmatildedynamicsDSSC}). This is the so called \textit{reduced dynamics} $\tau_{av}(t) \dot{\bar{u}}_0^{av} = -\bar{u}_0^{av}+u_{eq}$. Also replace $u_0$ by $u_{eq}$ in the predictor dynamics (\ref{eq:predictordynamicsDSSC}), leading to $\tau_m(t) \dot{\hat{\sigma}}=- \hat{\sigma} + \tau_m(t) k_o(t) [\tau_{av}(t) \dot{\bar{u}}_0^{av}]$. Since, during sliding mode at $\tilde{\sigma}=0$ one has $\hat{\sigma}=\sigma$, then one can further write
\begin{equation}
\tau_m(t)  \dot{\sigma}=- f_\sigma(\sigma) + \tau_m(t) k_o(t) [\tau_{av}(t) \dot{\bar{u}}_0^{av}]\,,
\label{eq:predictordynamicsDSSCproof2}
\end{equation}
leading to the time derivative of the synthesized  DSSC law $\bar{u}$ given by\footnote{The DSSC can  also be designed for high-order plants with arbitrary relative degrees, in this case, the synthesized  controller has more terms than (\ref{eq:equivcontroller2ndorderDSSC}).}  
\begin{eqnarray}
\dot{\bar{u}}(t)=-\dot{\bar{u}}_0^{av}(t) &=& -\left[\frac{1}{k_o(t) \tau_{av}(t)} \right]\dot{\sigma}(t)-\left[\frac{\sigma(t)}{k_o(t) \tau_{av}(t) \tau_m(t)} \right]\,. \label{eq:equivcontroller2ndorderDSSC}
\end{eqnarray}
Now, recalling that the dynamic functions $\tau_{av}$, $\tau_m$ and $k_o$ are, in fact, functions of $\sigma, e$ and $t$, then the terms in square brackets of  (\ref{eq:equivcontroller2ndorderDSSC}) are also functions of $\sigma, e$ and $t$. Therefore, if one can find appropriate functions $g_1(t)=g_1(\sigma(t),e(t),t)$ and $g_2(t)=g_2(\sigma(t),e(t),t)$  such that
\begin{equation}
\left[\frac{1}{k_o \tau_{av}} \right] = \frac{\partial [g_1 \sigma]}{\partial \sigma}  \,, \quad \mbox{and} \quad \left[\frac{\sigma}{k_o \tau_{av} \tau_m} \right]=\left[\frac{\partial (g_1 \sigma)}{\partial e} \dot{e}+\frac{\partial (g_1 \sigma)}{\partial t}+g_2 \sigma \right]\,,\label{eq:DSSCparXg1g2}
\end{equation}
then (\ref{eq:equivcontroller2ndorderDSSC})  can be rewritten as
\begin{equation}
    \dot{\bar{u}}(t)=-\frac{d\left[g_1(t) \sigma(t)\right]}{dt}-g_2(t) \sigma(t)\,.\label{eq:du}
\end{equation}
Therefore, by integrating both sides of (\ref{eq:du}), the synthesized DSSC law  can be written as:
\begin{equation}
\bar{u}(t)= -g_1(t) \sigma(t)-\int_{t_s}^t g_2(\tau) \sigma(\tau) d\tau + C_s\,, \quad \forall t \geq t_s\,, 
\label{eq:uDSSCsliding}
\end{equation}
where $C_s:=\bar{u}(t_s)+g_1(t_s) \sigma(t_s)$ is a constant\footnote{This constant is an unknown constant. However, this is not an issue since the expression for  $\bar{u}$ is used only for analysis purposes.} and $g_1(t)$ and $g_2(t)$ are nonlinear gains that should be designed so that the functions $k_o(t)$, $\tau_m(t)$ and $\tau_{av}(t)$ be positive, $\forall t \geq t_s$.
Depending on the choices for the nonlinear gains $g_1$ and $g_2$ the resulting synthesized controller has different structures and properties. To illustrate some of those possibilities, consider the examples that follow.

\section{Families of DSSC's Synthesized Controllers}

For simplicity, let $k_o \tau_{av}$, $\tau_m$, $g_1$ and $g_2$ being functions of $\sigma$, only. Thus,  $\frac{\partial g_1}{\partial e} =\frac{\partial g_1}{\partial t}=0$. 
In the following examples, we illustrate some particular controllers that can be synthesized from the DSSC: a PI controller with fixed gains and approximations for the standard STA and for the variable gain STA (VGSTA).

\begin{example}({\bf PI Control})
Choose $k_o$, $\tau_{av}$ and $\tau_m$ as positive constants in (\ref{eq:DSSCparXg1g2}). Then, from (\ref{eq:uDSSCsliding}), we arrive in $\bar{u}=u_{pi}+C_s$,  where $C_s:=\bar{u}(t_s)+ \sigma(t_s)/(k_o \tau_{av})$ and $u_{pi}$ is the PI control law
$$u_{pi}(t)= -g_1 \sigma(t)-g_2 \int_{t_s}^t \sigma(\tau) d\tau\,,$$
with proportional gain  $g_1=1/(k_o \tau_{av})$ and integral gain $g_2=1/(k_o \tau_{av} \tau_m)$. 
\end{example}

\begin{example}({\bf Standard STA})
\label{ex:STAstandard}
By setting 
$$g_1 \sigma := \kappa_1 \phi_1 :=\kappa_1 \sigma  |\sigma|^{-1/2} \quad \mbox{and} \quad g_2 \sigma := \kappa_2 \phi_2 :=\kappa_2 \phi_1 \phi_1^{'}=\frac{\kappa_2}{2} \mbox{sgn}(\sigma)\,,$$ 
where $\kappa_1$ and $\kappa_2$ are positive constant gains,  $\phi_1:= \sigma |\sigma|^{-1/2}$ and\footnote{Note that, $|\sigma|^{1/2} \mbox{sgn}(\sigma)=\phi_1$ and $\phi_1^{'}=|\sigma|^{-1/2} + \sigma \left[-\frac{1}{2} |\sigma|^{-3/2} \mbox{sgn}(\sigma)\right]=\frac{|\sigma|^{-1/2}}{2}$, when  $\sigma \neq 0$.} 
$2\phi_1^{'}=|\sigma|^{-1/2}$, and choosing in (\ref{eq:DSSCparXg1g2})
%
$$k_o \tau_{av}=\frac{1}{\kappa_1 \phi_1^{'}}=\frac{2}{\kappa_1} |\sigma|^{1/2}\,, \quad \mbox{and} \quad \tau_m=\frac{\kappa_1 \phi_1^{'} \sigma}{\kappa_2 \phi_2}=\frac{\kappa_1 \sigma}{\kappa_2 \phi_1}=\frac{\kappa_1}{\kappa_2} |\sigma |^{1/2}\,,$$ 
we arrive in $\bar{u}=u_{sta}+C_s$, where $C_s:=\bar{u}(t_s)+\kappa_1 \phi_1(t_s)$ and $u_{sta}$ is 
%
the standard STA law
\begin{equation}
u_{sta}(t):=-\kappa_1 |\sigma(t)|^{1/2} \mbox{sgn}(\sigma(t))-\frac{\kappa_2}{2} \int_{t_s}^t \mbox{sgn}(\sigma(\tau)) d\tau\,.
    \label{eq:defuSTAstd}
\end{equation}
%
However, since $k_o \tau_{av}$ and $\tau_m$ achieve zero at $\sigma=0$, the DSSC's averaging filter and predictor cannot be implemented, unless some approximation is made. The idea is to use approximations $\hat{\phi}_1$ and $\hat{\phi}_2$, for $\phi_1$ and $\phi_2$, respectively. \endproof
\end{example}

\begin{example}({\bf $\delta$-Approximation for the Standard STA: Case 1})\label{ex:STAdelta1}
In this example, we use approximations $\hat{\phi}_1$ and $\hat{\phi}_2$ for the functions $\phi_1$ and $\phi_2$ of the standard STA, respectively, to redefine $g_1 \sigma := \kappa_1 \hat{\phi}_1$ and $g_2 \sigma := \kappa_2 \hat{\phi}_2$, 
%
%
with positive constants $\kappa_1$ and $\kappa_2$,
so that we get $\bar{u}=\hat{u}_{sta}+C_s$, with $C_s:=\bar{u}(t_s)+\kappa_1 \hat{\phi}_1(t_s)$ and the following approximation for the standard STA law (\ref{eq:defuSTAstd})
\begin{equation}
\hat{u}_{sta}(t):= -\kappa_1 \hat{\phi}_1(t)-\kappa_2 \int_{t_s}^t  \hat{\phi}_2(\tau) d\tau\,,
    \label{eq:defuSTAstdapprox1}
\end{equation}
by choosing $k_o \tau_{av}=\frac{1}{\kappa_1 \hat{\phi}_1^{'}}$ and $\tau_m=\frac{\kappa_1 \hat{\phi}_1^{'} \sigma}{\kappa_2 \hat{\phi}_2}$ in (\ref{eq:DSSCparXg1g2}).
One possibility is to select $\hat{\phi}_1$ and $\hat{\phi}_2$ as the following $\delta$-approximations
\begin{equation}
\hat{\phi}_1:=\left[1-\frac{\delta \ln\left(\frac{|\sigma|^{1/2}+\delta}{\delta}\right)}{|\sigma|^{1/2}}\right] \phi_1\,, \quad \mbox{where} \quad \hat{\phi}_1^{'}=\frac{1}{2 (|\sigma|^{1/2}+\delta)}\,,
\label{eq:choicesSTAstandard2}
\end{equation}
and 
$$\hat{\phi}_2= \hat{\phi}_1 \hat{\phi}_1^{'}=\left[1-\frac{\delta \ln\left(\frac{|\sigma|^{1/2}+\delta}{\delta}\right)}{|\sigma|^{1/2}}\right] \left[\frac{|\sigma|^{1/2}}{2 (|\sigma|^{1/2}+\delta)} \right]\mbox{sgn}(\sigma)\,,$$ 
where $\delta>0$ is an arbitrarily small constant. This can be accomplished by selecting the DSSC's dynamic functions in (\ref{eq:DSSCparXg1g2}) as 
$$k_o \tau_{av}=\frac{2 }{\kappa_1} (|\sigma |^{1/2}+ \delta)\,,\quad \mbox{and} \quad \tau_m=\frac{\kappa_1}{\kappa_2} |\sigma |^{1/2} \left[1-\frac{\delta \ln\left(\frac{|\sigma|^{1/2}+\delta}{\delta}\right)}{|\sigma|^{1/2}}\right]^{-1}\,.$$
\endproof
\end{example}


\begin{example}({\bf $\delta$-Approximation for the Standard STA: Case 2}) \label{ex:STAdelta3}
Note that, for both choices in Example \ref{ex:STAdelta1}, 
the DSSC's dynamic functions are well defined for all finite $\sigma$, and one has that $\hat{\phi}_1 \rightarrow \phi_1=|\sigma|^{1/2} \mbox{sgn}(\sigma)$ and $\hat{\phi}_2= \hat{\phi}_1^{'} \hat{\phi}_1 \rightarrow  \phi_2=\phi_1^{'} \phi_1 = \mbox{sgn}(\sigma)/2$, as $\delta \rightarrow  0$. 
As an alternative, other choices for $\hat{\phi}_2$ can be explored, even not satisfying the relationship $\hat{\phi}_2 = \hat{\phi}_1^{'} \hat{\phi}_1$, for either of the two approximations for $\hat{\phi}_1$. For instance, 
$$\hat{\phi}_2=\frac{\sigma}{2(|\sigma|+\delta)}\,, \quad \mbox{or} \quad \hat{\phi}_2=\frac{\sigma}{2(|\sigma|^{1/2}+\delta)^2}\,.$$
In particular, for $\hat{\phi}_1=\left[1-\frac{\delta \ln\left(\frac{|\sigma|^{1/2}+\delta}{\delta}\right)}{|\sigma|^{1/2}}\right] \phi_1$ and $\hat{\phi}_2=\frac{\sigma}{2(|\sigma|^{1/2}+\delta)^2}$, one has that $k_o \tau_{av}=\frac{1}{\kappa_1 \hat{\phi}_1^{'}}$ and $\tau_m=\frac{\kappa_1 \hat{\phi}_1^{'} \sigma}{\kappa_2 \hat{\phi}_2}$ are given by
$$k_o \tau_{av}=\frac{2}{\kappa_1} (|\sigma |^{1/2}+ \delta)\,, \quad \mbox{and} \quad \tau_m=\frac{\kappa_1}{\kappa_2} (|\sigma |^{1/2}+ \delta)\,,$$
since $\hat{\phi}_1^{'}=\frac{1}{2 (|\sigma|^{1/2}+\delta)}$. In this case, the DSSC's dynamic functions are still well defined for all finite $\sigma$, $\hat{\phi}_1 \rightarrow \phi_1=|\sigma|^{1/2} \mbox{sgn}(\sigma)$ and $\hat{\phi}_2 \rightarrow  \phi_2=\phi_1^{'} \phi_1 = \mbox{sgn}(\sigma)/2$, as $\delta \rightarrow  0$, but $\hat{\phi}_2 \neq  \hat{\phi}_1^{'} \hat{\phi}_1$. \endproof  
\end{example}

\begin{example}({\bf $\delta$-Approximation for the Variable Gain STA})
The idea is similar to the standard STA approximation case. 
Now we redefine $g_1 \sigma:= \kappa_1 \hat{\phi}_1$ and $g_2 \sigma:= \kappa_2 \hat{\phi}_2$ 
%
%
with {\bf nonlinear variable gains} $\kappa_1(t)=\kappa_1(\sigma,e,t)>0$ and $ \kappa_2(t)=\kappa_2(\sigma,e,t)>0$ (with some abuse of notation) depending on the $\sigma$-dynamics (\ref{eq:sigmadynforsliding}) and the $e$-dynamics (\ref{eq:defedynamics}). This results in $\bar{u}=\hat{u}_{vgsta}+C_s$, where $C_s:=\bar{u}(t_s)+\kappa_1(t_s) \hat{\phi}_1(t_s)$ and
\begin{equation}
\hat{u}_{vgsta}(t)= -\kappa_1(t) \hat{\phi}_1(t)-\int_{t_s}^t \kappa_2(\tau) \hat{\phi}_2(\tau) d\tau\,,\label{eq:uDSSCslidingNEWgg}
\end{equation}
is a $\delta$-approximation for the  variable gain STA (VGSTA) control law \cite{morenoVGSTA1}, 
by choosing $k_o \tau_{av}$ and $\tau_m$ in (\ref{eq:DSSCparXg1g2}), appropriately. This is the focus of this paper and will be described in Section~\ref{sec:DSSC-VGSTA}.\endproof  
\end{example}


\subsection{Remarkable Features of the DSSC}

First, when compared with the variable gain STA (VGSTA) \cite{morenoVGSTA1, morenoVGSTA2} or  the standard STA \cite{morenoSTA1} \cite{morenoSTA2}, the synthesized DSSC can improve the robustness with respect to unmodelled dynamics. For the  plant (\ref{eq:plantx1})--(\ref{eq:plantx2}), this is due to the fact that the synthesized DSSC results in a $\delta$-approximation for the VGSTA (or standard STA). This approximation acts like a gain reducer near the origin of the error system's state space $(\sigma, e)$. 

From a theoretical point of view, this synthesized $\delta$-approximation becomes exactly the VGSTA (or the standard STA), as $\delta \rightarrow 0$. In addition, when the averaging filter's pass-band tends to infinity, the closed-loop dynamics with the synthesized DSSC law tends to be the closed-loop dynamics with the STA, in the absence of unmodelled dynamics, as described in the approximated analysis in Section~\ref{sec:undyn}. 

On the other hand, from a practical point of view, small values for $\delta$ are enough to obtain  similar results as the VGSTA (or the standard STA), far away from the origin, while assuring acceptable input disturbance rejection capabilities as the VGSTA (or the standard STA), near the origin.

Second, the initial value of the DSSC's control effort can start at zero by setting the averaging filter's initial condition at zero. In contrast, the VGSTA (or standard STA) control law can reach large values at $t=0$, unless an appropriate initialization is made.

\section{DSSC's Dynamic Functions Choice Related to the VGSTA}
\label{sec:DSSC-VGSTA}

Now, we explore one possible choice for the  DSSC dynamic functions that results, during sliding mode at $\tilde{\sigma}=0$, in a synthesized  equivalent controller approaching the VGSTA, far from the origin  of the state space $(\sigma, e)$, and acting like a reduced gain version of the VGSTA, near the origin. 

Henceforth, we consider $\hat{\phi}_2(\sigma):=\hat{\phi}_1(\sigma) \hat{\phi}_1^{'}(\sigma) $ with $\hat{\phi}_1$ defined as
\begin{equation}
 \hat{\phi}_1(\sigma):=\frac{\phi_a \sigma}{(|\sigma|^{1/2} + \delta)} + \phi_b \sigma =  \left[\frac{\phi_a |\sigma|^{1/2}}{(|\sigma|^{1/2} + \delta)}\right] |\sigma|^{1/2} \mbox{sgn}(\sigma)+ \phi_b \sigma\,,   \label{eq:defphi1}
\end{equation}
where $\delta>0$ is an arbitrary small constant and $\phi_a>0$ and $\phi_b>0$ are  design constants. Note that 
$$\hat{\phi}_1^{'}= \phi_a \left[\frac{(|\sigma|^{1/2}+2\delta)}{2(|\sigma|^{1/2}+\delta)^2}\right]+\phi_b\,.$$
By {\bf redefining} the nonlinear variable gains $g_1$ and $g_2$ as
$$g_1 \sigma=\kappa_1 \hat{\phi}_1\,, \quad \mbox{and} \quad g_2 \sigma =\kappa_2 \hat{\phi}_2 =\kappa_2 \hat{\phi}_1 \hat{\phi}_1^{'} \,, \quad \sigma \neq 0\,,$$
and $g_1=g_2=0$, for $\sigma=0$, 
%
%
%
%
%
%
%
the synthesized DSSC law (\ref{eq:uDSSCsliding}) becomes $\bar{u}=\hat{u}_{vgsta}+C_s$, with $C_s:=\bar{u}(t_s)+\kappa_1(t_s) \hat{\phi}_1(t_s)$ and
\begin{equation}
\hat{u}_{vgsta}(t)= -\kappa_1(t) \hat{\phi}_1(t)-\int_{t_s}^t \kappa_2(\tau) \hat{\phi}_2(\tau) d\tau\,.\label{eq:uDSSCslidingNEW}
\end{equation}
For $\phi_a=1$ and $\phi_b=\kappa_3$, the control law (\ref{eq:uDSSCslidingNEW}) is an approximation for the  VGSTA control law $u_{vgsta}(t):=-\kappa_1 \phi_1(\sigma(t)) - \int_{t_s}^t \kappa_2 \phi_2(\sigma(\tau)) d\tau$,  of \cite{morenoVGSTA1, morenoVGSTA2}, with $\phi_1(\sigma):=|\sigma|^{1/2} \mbox{sgn}(\sigma)+\kappa_3 \sigma$ and $\phi_2(\sigma):=\phi_1^{'}(\sigma) \phi_1(\sigma)$. 
%

Now, since the variable gains $\kappa_1>0$ and $ \kappa_2>0$ are functions of the time $t$ and the plant's states $\sigma$ and $e$, then from  (\ref{eq:DSSCparXg1g2}), one has to select DSSC' s dynamic functions $k_o$, $\tau_{av}$ and $\tau_m$  to satisfy
$$\left[\frac{1}{k_o \tau_{av}} \right] =\left[\kappa_1^{'} \hat{\phi}_1 + \kappa_1 \hat{\phi}_1^{'}  \right]\,, \quad \mbox{and} \quad \left[\frac{\sigma}{k_o \tau_{av} \tau_m} \right]=\hat{\phi}_1 \left[\frac{\partial \kappa_1}{\partial e} \dot{e}+\frac{\partial \kappa_1}{\partial t}+\kappa_2 \hat{\phi}_1^{'}\right]\,.$$
%
%
%
Thus, one has to select $k_o$, $\tau_{av}$ and $\tau_m$ so that
\begin{equation}
k_o \tau_{av}:=\frac{1}{\left[\kappa_1^{'} \hat{\phi}_1 + \kappa_1 \hat{\phi}_1^{'}  \right]}\,,
    \label{eq:defkotauav}
\end{equation}
and
\begin{equation}
\tau_m:=\frac{\left[\kappa_1^{'} \hat{\phi}_1 + \kappa_1 \hat{\phi}_1^{'}  \right]}{\left[\frac{\phi_a}{(|\sigma|^{1/2} + \delta)} + \phi_b\right] \left[\frac{\partial \kappa_1}{\partial e} [-l_0 e+\sigma]+\frac{\partial \kappa_1}{\partial t}+\kappa_2 \hat{\phi}_1^{'}\right]}\,,
    \label{eq:deftaum}
\end{equation}
where the relationship $\dot{e}=-l_0 e+\sigma$ was used. Note that, in (\ref{eq:defkotauav}), 
an extra degree of freedom is allowed for choosing $k_o$ and $\tau_{av}$: (i) $k_o$ being a constant function and $\tau_{av}$ time-varying function or vice-versa and (ii) both being time-varying functions.

In fact, the variable gains $\kappa_1(\sigma,e,t)>0$ and $ \kappa_2(\sigma,e,t)>0$ must be designed so that $\tau_{av}>0$ and $\tau_m>0$ are well-defined for all finite values of $\sigma, e$. 
The DSSC's dynamic functions and the design guidelines of the corresponding control parameters are summarized in Table~\ref{tab:par}.

\begin{table}[!ht]
    \centering
    \begin{tabular}{c|c}
        \multicolumn{2}{c}{\parbox{14cm}{The DSSC algorithm is composed by: the tracking error in (\ref{eq:defe}), the relative degree one variable $\sigma$ in (\ref{eq:defsigma}), predictor in (\ref{eq:predictordynamicsDSSC}) with the discontinuous  term in (\ref{eq:defu0DSSC}),  modulation function in  (\ref{eq:defvarrho}), sliding variable $\tilde{\sigma}$ in (\ref{eq:defsigmatildeDSSC}),  DSSC law in (\ref{eq:defDSSC}), complete control in (\ref{eq:updef}) and smooth averaging filter in (\ref{eq:u0avdynamicsDSSC}).}}\\
        \hline \\
        Dynamic Functions & Dynamic Functions (Cont.)\\
        \hline \\
        $k_o \tau_{av}:=\frac{1}{\left[\kappa_1^{'} \hat{\phi}_1 + \kappa_1 \hat{\phi}_1^{'}  \right]}$ & $\tau_m:=\frac{\left[\kappa_1^{'} \hat{\phi}_1 + \kappa_1 \hat{\phi}_1^{'}  \right]}{\left[\frac{\phi_a}{(|\sigma|^{1/2} + \delta)} + \phi_b\right] \left[\frac{\partial \kappa_1}{\partial e} [-l_0 e+\sigma]+\frac{\partial \kappa_1}{\partial t}+\kappa_2 \hat{\phi}_1^{'}\right]}$\\
        $\hat{\phi}_1(\sigma):=\frac{\phi_a \sigma}{(|\sigma|^{1/2} + \delta)} + \phi_b \sigma$ & $\hat{\phi}_1^{'} =  \phi_a \left[\frac{(|\sigma|^{1/2}+2\delta)}{2(|\sigma|^{1/2}+\delta)^2}\right]+\phi_b$\\
        $\kappa_1:= (\kappa_{a}|\sigma| + \kappa_{b}|e|+\kappa_{c})^2+\kappa_{d}$ & $\kappa_2=2\epsilon \kappa_1+\gamma$\\
        \hline \\
        Design Inequalities & Control Parameters\\
        \hline \\
        %
        $4 \epsilon k_p (\gamma k_p-4\epsilon^2) >1$ & $\gamma:=\frac{(1+\varepsilon_1)}{4 \epsilon \underline{k}_p^2}+\frac{4 \epsilon^2}{\underline{k}_p}$  \\
        %
        $\phi_b > \frac{l_0}{\epsilon}$ &  $\phi_b:=\frac{l_0}{\epsilon}+\varepsilon_3$ \\
        $\kappa_c > \max\left\{\frac{(8 \epsilon^2+ 2 \gamma k_p)}{[4 \epsilon k_p (\gamma k_p-4\epsilon^2)-1]}\,, \frac{(\bar{k}_{d3} \phi_b + k_\sigma)}{\phi_b^2}\right\}$ &  $\kappa_c := \max\left\{\frac{(8 \epsilon^2+ 2 \gamma \bar{k}_p)}{\varepsilon_1}\,, \frac{(\bar{k}_{d3} \phi_b + k_\sigma)}{\phi_b^2}\right\}$\\
        $\kappa_{b}>\frac{(\bar{k}_{d1} \phi_b + \bar{k}_p c_{ie})}{\phi_b^2}$ & $\kappa_{b}:=\left[\frac{(\bar{k}_{d1} \phi_b + \bar{k}_p c_{ie})}{\phi_b^2}\right]+\varepsilon_2$\\
        $\kappa_{a}>\max\left\{\frac{(\bar{k}_{d2} \phi_b + \bar{k}_p c_{i\sigma})}{\phi_b^2}\,, \frac{\kappa_b}{l_0}\right\}$ & $\kappa_{a}:=\left[\max\left\{\frac{(\bar{k}_{d2} \phi_b + \bar{k}_p c_{i\sigma})}{\phi_b^2}\,, \frac{\kappa_b}{l_0}\right\}\right]+\varepsilon_2$\\
        $\kappa_d > \frac{(8 \epsilon^2 \gamma k_p + 4 \epsilon^2)}{4 \epsilon k_p (\gamma k_p-4\epsilon^2) }$ & $\kappa_d:=\frac{(8 \epsilon^2 \gamma k_p + 4 \epsilon^2)}{4 \epsilon k_p (\gamma k_p-4\epsilon^2) }$\\
         \hline
   \end{tabular}
    \caption{DSSC's dynamic functions and parameters. The free parameters are: $ l_0, \phi_a, \epsilon, \delta >0$, and $\varepsilon_i>0$ ($i=1,2,3$).}
    \label{tab:par}
\end{table}

\subsection{Closed Loop Convergence Results}\label{sec:stability}

Before state the main theorem, let define the \textit{nominal control law} composed by: (i) a feedforward term $u_m^n(t)$; (ii)  $u_p^n(e(t))$, representing a proportional feedback action; (iii) $u_d^n(\sigma(t))$, contributing to a derivative plus proportional feedback action (since $\sigma=l_0 e+\dot{e}$); and (iv) $u_i^n(t)=\int_0^t \bar{u}^n_i(e(\tau),\sigma(\tau)) d\tau$, as an integral feedback action. The nominal control is written in the form  
\begin{equation}
u^n(t):= u_p^n(t)+u_d^n(t)+u_i^n(t)+u_m^n(t)\,,
%
\label{eq:defunomGERAL}
\end{equation}
where $u_p^n(t)=u_p^n(e(t))$ and $u_d^n(t)=u_d^n(\sigma(t))$. 
For simplicity and without loss of generality, we restrict the nominal control to have terms that satisfy the following additional assumption:
\begin{description}
\item[{\bf (A4)}] There exist  non-negative constants $c_\sigma$, $c_{e1}$, $c_{e2}$, $c_{i\sigma}$, $c_{ie}$ and $c_{m}$ such that
\begin{eqnarray}
    |u_p^n(e)| &\leq& c_{e1} |e|\,, \quad \left|\frac{d u_p^n(e)}{d e}\right| \leq c_{e2}\,, \quad |u_d^n(\sigma)| \leq c_\sigma |\sigma|\,, \nonumber \\ 
    |\bar{u}_i^n(e,\sigma)| &\leq&  (c_{i\sigma} |\sigma|+c_{ie}|e|)|\sigma|\,, \quad \mbox{and} \quad |u_m^n| \leq c_m\,.\label{eq:unombounds}
\end{eqnarray}
%
\end{description}
It must be highlighted that the nominal control 
is not regarded as a disturbance and can be disregarded when the plant uncertainty is large. 
\begin{rmk}
One particular choice for the nominal control is given by $u_{d}^n(\sigma) =-c_\sigma \sigma$, $u_p^n(e)=- c_e e$, $u_m^n=- c_{m1} \dot{y}_m - c_{m2} \ddot{y}_m$ and $\bar{u}_i^n=0$, leading to the linear nominal control
\begin{equation}
u^n:=- c_e e -c_\sigma \sigma  - c_{m1} \dot{y}_m - c_{m2} \ddot{y}_m\,,
\label{eq:defunom}
\end{equation}
with $c_\sigma, c_e, c_{m1}, c_{m2}$ being uniformly norm bounded signals (in general, constants) designed for stabilization and/or to take advantage of some nominal knowledge of the plant.\endproof
\end{rmk}

The main results are summarized in the following theorem.

\begin{thm}\label{theorem:DSSC}
Consider the  plant represented in (\ref{eq:planteta})--(\ref{eq:plantoutput}),  Assumptions ({\bf A0})--({\bf A4}) and the DSSC' s algorithm and parameters described in Table~\ref{tab:par},  with the gain 
$\phi_b>0$. Then, for $\phi_b$ sufficiently large, the output tracking error is globally exponentially convergent w.r.t. a small residual set of order $\mathcal{O}(1/\phi_b^2)$, satisfying the inequality
\begin{equation}
|e(t)| \leq \mathcal{O}(1/\phi_b^2) + \pi_e\,,
\label{eq:boundone}
\end{equation}
where $\pi_e$ is an exponentially decaying term depending on the initial conditions and this residual set does not depend on the initial conditions. 
In addition, all closed-loop signals remain uniformly bounded, finite-time escape is avoided and the sliding variable becomes identically null after some finite time $t_s \geq 0$. 
\end{thm}
\proof 
For this particular case, where 
the functions $\tau_{av}(t)$, $\tau_m(t)$ and $k_o(t)$ are chosen according to (\ref{eq:defkotauav}) and  (\ref{eq:deftaum}), an approximation for VGSTA is synthesized during the sliding mode, and the main idea of the proof is as follows. Firstly, we prove that finite-time escape cannot occur before $\tilde{\sigma}(t)=0$, i.e., before sliding mode takes place. Secondly, once $\tilde{\sigma}=0$ enters in sliding motion in finite time, then  the proof follows the general approach of the VGSTA's convergence proof given in \cite{derafa2012super}, \cite{morenoVGSTA1} and  \cite{morenoVGSTA2}. The main difference is the introduction of the Small-Gain Theorem to deal with the $\delta$-approximation of the VGSTA. See the \ref{ap:ProofTheorem} for the complete proof. The case of general choices for $\tau_{av}(t)$, $\tau_m(t)$, and $k_o(t)$ was left for a future work.\endproof

\begin{rmk}({\bf Local or Semi-Global Results When $\phi_b=0$})
For the case $\phi_b=0$, one can perform the stability analysis for the case that the DSSC's synthesized  law results in the approximation for the standard STA, which is conducted in a similar manner, as in Theorem~\ref{theorem:DSSC}. The main difference is that the gains $\kappa_1$ and $\kappa_2$ can be designed constant for local/semi-global results around the origin of ($\sigma, e$). The formulation and analysis for this case are omitted to save space.\endproof
\end{rmk}

\begin{rmk}({\bf Regulation Mode Case})
In the regulation mode, one has that $\bar{\beta}_m$ and $\dot{y}_m$ are zero in the small-gain based analysis, 
since $k_{d5}=0$ and $\alpha_{d2}=\alpha_{d3}=0$ when a constant disturbance is under consideration. Thus, in this case, the tracking error $e$ converges to zero, exponentially, and a constant disturbance is totally rejected.\endproof
\end{rmk}

\begin{rmk}({\bf Prescribed Finite-Time Convergence}) Additionally to Theorem~\ref{theorem:DSSC}, prescribed finite-time convergence for the residual set can be assessed. 
The term $-2 \epsilon \frac{1}{\mu}\frac{V}{\lambda_{max}\{P\}}$, in (\ref{eq:dV1}), is responsible to assure that $\sigma$ and the tracking error $e$ both reach a residual set in a prescribed finite-time. This analysis is left for future work to save space. \endproof
\end{rmk}

\subsection{Robusteness w.r.t. Unmodelled Dynamics: Approximated Analysis}\label{sec:undyn}

Assume that an unmodelled dynamic represented by a transfer function of the form 
$$G_\mu(\mu s):=1+\mu s W_\mu(\mu s)\,,$$
where $W_\mu(\mu s)$ is stable and strictly proper, is now in series with the plant input $u_p=u+u^n$, in (\ref{eq:plantx2}), i.e. 
$$u_p=u+u^n+d_\mu\,, \quad d_\mu:=\mu s W_\mu(\mu s) (u+u^n)\,,$$
\textit{modulo} exponentially decaying terms due to the unmodelled dynamics initial conditions. This extra term $d_\mu$ can be regarded as an additional input disturbance and {\bf incorporated} in the input disturbance $d$, in (\ref{eq:plantx2}).
%
Recalling that $u=-u_{0}^{av}$ and $\tau_{av} \dot{u}_{0}^{av}=-u_{0}^{av}+u_0$, then one can write
$$d_\mu:=\mu s W_\mu(\mu s) u=\mu W_\mu(\mu s) \dot{u} = \frac{\mu}{\tau_{av}} W_\mu(\mu s) (u_{0}^{av}-u_0)\,.$$
As some examples for the unmodelled dynamics transfer function, one has: (i) $W_\mu(\mu s)=-\frac{1}{(\mu s+1)}$ and $G_\mu(\mu s)=\frac{1}{(\mu s+1)}$; and (ii) $W_\mu(\mu s)=-\frac{(\mu s+2)}{(\mu s+1)^2}$ and $G_\mu(\mu s)=\frac{1}{(\mu s+1)^2}$. 
Thus, this additional disturbance $d_\mu$ is a filtered version of the averaging control $u_0^{av}$ and the discontinuous control $u_0$, via a proper and stable transfer function of order $\mathcal{O}(\mu/\tau_{av})$. Thus, for $\mu/\tau_{av}$ sufficiently small and
despite some parasitic dynamics $\mu$, the ideal sliding mode can still be enforced after some finite time, for the appropriate design of the modulation function.

In order to explain the main idea, for simplicity, consider that: (i) $u^n=0$, (ii) $\tau_{av}$ is a constant and $\tau_m= (|\sigma |^{1/2}+ \delta)/\kappa_1$, with $\kappa_1$ and $\delta$ being positive constants, (iii) the system has order two (no zero dynamics) and is perfectly known ($a_p^n=a_p$ and $k_p^n=k_p$), and (iv) the nominal control is given by $k_p^n u^n:= -(l_0-a_p^n) \dot{y}+\dot{\sigma}_m$. So that $d_\sigma$ and the $\sigma$-dynamics, both in (\ref{eq:sigmadynforsliding}), become
\begin{equation}
d_\sigma:=k_p^n d \quad \mbox{and} \quad \dot{\sigma}=k_p^n (-u_{0}^{av}+d)\,,\label{eq:defdsigmaPerfectandsigmadyn}
\end{equation}
respectively, where we have replaced $u$ by the DSSC's control law $u=-u_0^{av}$, with $u_0^{av}$ in (\ref{eq:u0avdynamicsDSSC}).

%
Now, one can subsequently conclude that: (i) the disturbance term $\tilde{d}_1$, in (\ref{eq:defdtil1}), reduces to $\tilde{d}_1= (-l_0 \dot{y}+ \dot{\sigma}_m +k_p^n d)$; (ii) $\tilde{d}_2$, in (\ref{eq:defdtil2}), reduces to $\tilde{d}_2= (\sigma/\tau_m+l_0 \dot{y}- \dot{\sigma}_m)$; (iii) $\tilde{d}_1+\tilde{d}_2= k_p^n d+ \sigma/\tau_m$; (iv) and $d_0$, in (\ref{eq:defd0}), reduces to 
$d_0 =(k_o-k_p^n)u_0^{av}+ k_p^n d+ \sigma/\tau_m$.
%
%
Let $\bar{u}_0^{av}=u_0^{av}$ be the solution of the \textit{reduced dynamics}, resulting by replacing  the discontinuous control $u_0$ in (\ref{eq:u0avdynamicsDSSC}) by the following \textit{equivalent control} $u_{eq}=d_0/k_o$, obtained from the $\tilde{\sigma}$-dynamics (\ref{eq:sigmatildedynamicsDSSC}). Then, one can write
$$u_{eq}=\frac{d_0}{k_o} = \left(1-\frac{k_p^n}{k_o}\right)\bar{u}_0^{av}+\frac{k_p^n}{k_o} d+\frac{1}{k_o \tau_m} \sigma\,.$$
Now, an {\bf approximated analysis} can be carried out for understanding the superior performance of the DSSC in comparison to the STA, in the presence of unmodelled dynamics. Since the averaging control $\bar{u}_0^{av}$ is an approximation of the equivalent control $u_{eq}$, for $\tau_{av}$ sufficiently small \cite{utkin1992scope}, one has that $u_{eq}\approx \bar{u}_0^{av}$ implies
%
%
$$\bar{u}_0^{av}\approx d+\frac{1}{k_p^n \tau_m} \sigma\,.$$
With $u_0^{av}=\bar{u}_0^{av}$ in   
(\ref{eq:defdsigmaPerfectandsigmadyn}), the closed-loop $\sigma$-dynamics can be approximated by 
$$\dot{\sigma} \approx -\frac{1}{\tau_m}  \sigma \approx -\frac{\kappa_1}{(|\sigma|^{1/2}+\delta)}  \sigma \approx \kappa_1 |\sigma|^{1/2} \mbox{sgn}(\sigma)\,,$$
for $\delta>0$ and small. Thus, the closed-loop $\sigma$-dynamics with the DSSC law approaches the closed-loop $\sigma$-dynamics with the standard STA, without the presence of input disturbance (with $\kappa_2=0$). Finally, since $d$ incorporates the equivalent disturbance $d_\mu$ generated by the unmodelled dynamics, it becomes evident that the DSSC should outperform the corresponding STA.


\section{Application: Trajectory Tracking of UAVs}
\label{sec:UAV}





In this section, we presented the UAV's dynamic model, including the aerodynamic effects, and the corresponding dynamic version for pitch and roll small angles which is used to design the inner velocity control loops. 

We consider the UAV application based on the following  premises and motivations:
\begin{enumerate}
\item Low/medium velocity profiles are considered so that the motors and the motors drivers (ESCs) can be neglected, as well as, the internal Kalman Filter dynamics of the UAV leading to the availability of the full UAV state vector.

\item The proposed DSSC scheme acts as an outer controller which provides velocity commands as references for the inner control loops. These inner controls are designed base on an UAV's dynamic model for pitch and roll small angles and all controllers are applied to the full UAV's dynamic model, including the aerodynamic effects. 

\item The inner control was developed as simply as possible to be representative of the unavailable internal control loops in the DJI M600, without putting any effort into stability analysis or tuning control parameters methodologies. The consistency of the inner control loops developed here was verified first with the DJI Assistant $2$ Simulator\footnote{The  DJI Assistant $2$ Simulator\footnote{https://www.dji.com/downloads/softwares/assistant-dji-2.  Notice that the DJI Assistant $2$ Simulator does not provide access to the inner control loops.}, a program developed by DJI company that allows the users to upload flight data, calibrate vision sensors, and provide a simulator with dynamics very close to the real DJI M600 drone.} and then with experimental data.

\item The inertia tensor of each propeller hub (propeller plus motor) and the inertia tensor of the UAV's structure are constant diagonal matrices.

\end{enumerate}




\subsection{The UAV's Dynamic Model with Aerodynamics Effects }
\label{sec:dronemodel}

In this section, the UAV's dynamic model is developed for low-velocity profiles. It means that  the  dynamics of the motors and the motors' drivers (ESC’s) can be neglected, while the more relevant effects are due to the aerodynamic forces and torques. 

By using the well-known
Newton-Euler method, the UAV's dynamics can be written as
\begin{eqnarray}
\mathcal{M} \dot{v}&=&-\mathcal{M} g e_3+ R f e_3 +F_{drag}\,, \nonumber  \\
\mathcal{J} \dot{\Omega}&=& -\left(\Omega \times \mathcal{J} \Omega\right) + M+ \tau_{drag} + \tau_{dist}\,, \label{eq:fullUAVmodel}\\ 
\dot{R} &=& R \hat{\Omega}\,,\nonumber
\end{eqnarray}
%
where $\Omega=\left[\begin{array}{ccc}\Omega_x & \Omega_y &\Omega_z\end{array}\right]^T \in \re^{3}$  is the UAV's angular velocity represented in the body frame, $v=\left[\begin{array}{ccc}v_x & v_y &v_z\end{array}\right]^T \in \re^{3}$ is the UAV's mass center linear velocity vector (in the inertial frame), $R$ represents the rotation matrix from the body frame to the inertial frame, $g$ is the gravity acceleration and  $n_r$ represents the number of identical rotors and propellers located at the vertices of a polygonal,
%
$\tau_{dist}:=\left[\begin{array}{ccc} I_{z}  \Omega_y \sum_{i=1}^{n_r} \dot{\theta}_i & 
-I_{z}   \Omega_x  \sum_{i=1}^{n_r} \dot{\theta}_i &
I_{z}  \sum_{i=1}^{n_r}\ddot{\theta}_i\end{array}\right]^T$,
%
%
%
%
and $\mathcal{J}=\mbox{diag}\left(\left[\begin{array}{ccc}n_r I_{xy} +I_{bx} & n_r I_{xy} +I_{by}  & n_r I_z +I_{bz}\end{array}\right]\right)=\mbox{diag}\left(\left[\begin{array}{ccc}\mathcal{J}_{x} & \mathcal{J}_{y} & \mathcal{J}_{z}\end{array}\right]\right)$.  Let $i=1,\ldots,n_r$.
The UAV's dynamics was implemented assuming that: (i) the {\bf constant} inertia tensor of the $i$-th propeller hub (propeller plus motor), represented in $\mathcal{E}_i$ (the $i$-th propeller frame), is a diagonal matrix $I_i=\mbox{diag}\left(\left[\begin{array}{ccc}I_{xi} & I_{yi} & I_{zi}\end{array}\right]\right)$, with, $I_{xi}=I_{yi}=I_{xy}$ and $I_{zi}=I_z$ ($\forall i$); and the {\bf constant} inertia tensor of the UAV's structure, represented in $\mathcal{E}_b$, is a diagonal matrix $I_b=\mbox{diag}\left(\left[\begin{array}{ccc}I_{bx} & I_{by} & I_{bz}\end{array}\right]\right)$. Moreover, we consider  $m_1=m_2=\ldots=m_{n_r}=\bar{m}$ and 
$\mathcal{M}:=m+n_r \bar{m}$, $k_{T_1}=k_{T_2}=\ldots=k_{T_{n_r}}=k_{T}$.

The following terms  incorporate all drag effects on the UAV: 
$$F_{drag}:=\sum_{i=1}^{n_r} F_{d i}  + F_d\,, \quad \mbox{and} \quad \tau_{drag}:=\sum_{i=1}^{n_r} (p_{bi} \times R^T F_{d i})\,,$$
where $p_{bi}$ is the position vector of the origin of the propeller frame relative to the origin of the body frame (represented in the body frame), and for the control allocation, 
$M:=\sum_{i=1}^{n_r} (p_{bi} \times T_i) +\sum_{i=1}^{n_r} \tau_{di}$ is the \textit{net moment}
and 
$f = \sum_{i=1}^{n_r} f_i$ is the \textit{net thrust magnitude}. The drag terms $F_{di}$ and  $F_d$ are defined in what follows.
To build a reliable simulation, we consider that the aerodynamic forces and torques acting on the $i$-th propeller and on the UAV's structure are: 
(i) the {\bf Propeller Aerodynamic Thrust} $T_i$ (body frame) with the thrust magnitude $f_i:=k_{T_i} \dot{\theta}_i^2$  proportional to the rotor  spin rate square via Rayleigh's equation,  where $k_{T_i}>0$ is the thrust aerodynamic constant and $\dot{\theta}_i$ is the $i$-th propeller spin rate; 
(ii) the {\bf Propeller Aerodynamic Drag Torque} $\tau_{d i}$ (body frame), with magnitude $|\tau_{d i}|=c_\tau k_{T_i} \dot{\theta}_i^2$, torque direction $s_i=\mbox{sgn}(\tau_{d i})=-\mbox{sgn}(\dot{\theta}_i)$ and aerodynamic torque constant $c_\tau > 0$; 
(iii) the {\bf Propeller Aerodynamic Drag Force} $F_{d i}:=- K_{F_{di}} \, |\dot{\theta}_i|\, v_{r i}$ (inertial frame), where $v_{ri}:=v_i-v_w$ is the propeller air-relative velocity, $v_i$ is the linear velocity of the $i$-th propeller frame, $v_w$ is the wind velocity, both represented in the inertial frame and $K_{F_{di}}>0$ is the propeller aerodynamic drag force matrix coefficient; (iv) the {\bf UAV Aerodynamics Drag Force on the Structure} $F_d:=- R K_{F_{d}}R^T v_r \,  \|v_r\|$, where $v_{r}:=v-v_w$ is  the air-relative velocity,  and $K_{F_{d}}>0$ is the structure aerodynamic drag force matrix coefficient.

\subsection{Inner Controllers}\label{sec:InnerControls} 

For the inner control loops design, $F_{drag}$, $\tau_{drag}$ and $\tau_{dist}$ are considered as disturbances. The Propeller Aerodynamic Thrust ($T_i$) and Propeller Aerodynamic Drag Torque  ($\tau_{d i}$) provide the control effort\footnote{As mentioned before, for low-velocity profiles, the motors and the motors' drivers (ESC’s) can be neglected.} via $M$ and $f$.

Now, for design the inner control laws, we consider an {\bf approximated} dynamics obtained by assuming small pitch and roll angle variation ($\theta,\phi$), so that the  Jacobian of the representation can be approximated by the identity matrix ($J_R \approx I$), leading to the approximation for the angular velocity $\Omega \approx \left[\begin{array}{ccc}\dot{\phi} & \dot{\theta} & \dot{\psi}\end{array}\right]^T$, where $\psi$ is the yaw angle.
The approximated dynamics is given by 
\begin{eqnarray}
\mathcal{M} \dot{v}_x &=& (\sin{\psi}\sin{\phi}+\cos{\phi} \sin{\theta} \cos{\psi})  f +\mathcal{D}_x\,, \nonumber  \\
\mathcal{M} \dot{v}_y &=& (-\cos{\psi}\sin{\phi}+\sin{\psi} \sin{\theta} \cos{\phi})  f+\mathcal{D}_y\,, \nonumber \\
\mathcal{M} \dot{v}_z &=& -\mathcal{M}  g + (\cos{\phi} \cos{\theta})  f+\mathcal{D}_z\,, \nonumber \\
\ddot{\phi}&=& \dot{\theta} \dot{\psi} \left(\frac{\mathcal{J}_y-\mathcal{J}_z}{\mathcal{J}_x}\right) + \frac{M_x}{\mathcal{J}_x} + \frac{\mathcal{D}_{\Omega_x}}{\mathcal{J}_x}\,, \nonumber \\
\ddot{\theta}&=& \dot{\phi} \dot{\psi} \left(\frac{\mathcal{J}_z-\mathcal{J}_x}{\mathcal{J}_y}\right) + \frac{M_y}{\mathcal{J}_y}+\frac{\mathcal{D}_{\Omega_y}}{\mathcal{J}_y}\,, \nonumber \\
\ddot{\psi} &=& \dot{\theta} \dot{\phi} \left(\frac{\mathcal{J}_x-\mathcal{J}_y}{\mathcal{J}_z}\right) + \frac{M_z}{\mathcal{J}_z}+\frac{\mathcal{D}_{\Omega_z}}{\mathcal{J}_z}\,, \nonumber
\end{eqnarray}
with $M=\left[\begin{array}{ccc}M_x & M_y &M_z\end{array}\right]^T$, and  
disturbances
$\left[\begin{array}{ccc}\mathcal{D}_x & \mathcal{D}_y & \mathcal{D}_z\end{array}\right]^T=F_{drag}^T$, 
and
$\left[\begin{array}{ccc}\mathcal{D}_{\Omega_x} & \mathcal{D}_{\Omega_y} & \mathcal{D}_{\Omega_z}\end{array}\right]^T=\tau_{drag}^T + \tau_{dist}^T$.
The inner controllers ensure that the UAV is in a velocity-controlled flight mode so that almost global asymptotic velocity tracking is assured in the sense that the  velocity vector $v(t)$ of the UAV asymptotically tracks a commanded velocity vector $v_d(t)$, i.e., 
$$v(t)=\left[\begin{array}{c}
v_x(t)\\
v_y(t)\\
v_z(t)
\end{array}\right]=v(t) \rightarrow v_d(t):=\left[\begin{array}{c}
u_x(t)\\
u_y(t)\\
u_z(t)
\end{array}\right]\,, \ \mbox{as} \ t \rightarrow \infty\,.$$ 
Moreover, we also assume that in this flight mode the yaw angle rate also asymptotically tracks a commanded yaw velocity: 
$$\dot{\psi}(t) \rightarrow \dot{\psi}_d(t) :=u_\psi(t)\,, \quad as \quad t \rightarrow \infty\,.$$

In what follows, we only describe the inner controller for the altitude. The same idea is employed in the other degrees of freedom, but it is  omitted to save space. 

\medskip
{\bf Altitude Control ($z$)}
\medskip

The {\bf altitude dynamics} can be expressed as
\begin{equation}
\dot{v}_z = \frac{k_h(t)}{\mathcal{M}}f - g +\frac{\mathcal{D}_z}{\mathcal{M}}\,,\label{eq:dynZ_kh_kg_d}
\end{equation}
with $k_h(t):=\cos(\phi(t) \cos(\theta(t))$ and $f(t)$ being the control variable. Notice that  $\phi(t)$  and $\theta(t)$ can be treated as exogenous available signals. The altitude control law is given 
$$f=(\mathcal{U}_z+g) \mathcal{M}/k_h\,,$$
which is composed of a feedback linearization term (which is parameter dependent) plus the PI-control law 
%
$$\mathcal{U}_z:= -k_d^z (v_z-u_{z})-k_i^z \int_0^t (v_z(\tau)-u_{z}(\tau)) d\tau\,,$$
leading to the second-order closed-loop dynamics
\begin{equation}
\dot{v}_z +k_p^z (v_z-u_z) + k_i^z \int_0^t (v_z(\tau)-u_{z}(\tau)) d\tau=\frac{\mathcal{D}_z}{\mathcal{M}}\,,\label{eq:closedloopzVelINT1}
\end{equation}
when $\mathcal{M}$ and $g$ are perfectly known and $k_h$ is perfectly cancelled. This results in a relative degree one closed-loop dynamics from the velocity command input $u_z$ to the actual UAV velocity $v_z$, due to the proportional control action.
Letting $e_z:=v_z-u_z$, one can write
%
\begin{equation}
\ddot{e}_z +k_p^z \dot{e}_z + k_i^z e_z=\frac{\dot{\mathcal{D}}_z}{\mathcal{M}}-\ddot{u}_z\,,\label{eq:closedloopzVelINT2}
\end{equation}
leading to conclude that, for low acceleration commands ($\ddot{u}_z \approx 0$) and for low aerodynamic drag ($\mathcal{D}_z \approx 0$), one has $e_z(t)$ approaching zero, as $t \rightarrow \infty$, for appropriate choices for the control gains $k_p^z>0$ and $k_i^z>0$.

With this inner control scheme, the closed-loop  dynamic behavior from the velocity command input $u_z$ to the actual UAV velocity $v_z$ is given by
\begin{equation}
\ddot{v}_z = -k_p^z (\dot{v}_z-\dot{u}_z) - k_i^z (v_z-u_{z}) +\frac{\dot{\mathcal{D}}_z}{\mathcal{M}}\,,\label{eq:closedloopzuzTovz}
\end{equation}
which can be represented by a {\bf relative degree one and minimum phase} system in the normal form
\begin{eqnarray}
\dot{\eta}_z&=&-\left[\frac{k_i^z}{k_p^z}\right] \eta_z + v_z\,, \label{eq:etadynamicsz} \\
\dot{v}_z &=& \left[\frac{k_i^z-(k_p^z)^2}{k_p^z}\right]  v_z - \left[\frac{(k_i^z)^2}{(k_p^z)^2}\right]  \eta_z + k_p^z (u_z+d_z)\,,\label{eq:vdynamicsz}
\end{eqnarray}
by transforming the state vector 
$\left[\begin{array}{cc}\dot{v}_z & v_z\end{array}\right]^T$ to $\left[\begin{array}{cc}\eta_z & v_z\end{array}\right]^T$, where the zeros dynamics state vector $\eta_z \in \re$ is given by 
$\eta_z:=k_p (k_i-k_p^2) v_z/k_i^2 - k_p^2 (\dot{v}_z-k_p u)/k_i^2$. The disturbance $d_z(t)$ can incorporate the disturbance $\frac{\mathcal{D}_z}{k_p^z \mathcal{M}}$ only, or other eventually remaining terms due to any mismatch parameters in the feedback linearization control term. 

Exactly tracking in the inner velocity control loop is not needed since the outer position control loop can compensate for these uncertainties. Moreover, a more elaborate inner controller could be considered \cite{lee2010control}, but this is not the focus of this paper, and this simple feedback linearization plus PI control strategy has provided consistent results with the DJI Assistant $2$ simulator and with  the experimental data obtained with the DJI M600.

\begin{rmk}{\bf (First-Order System)}
Note that, when $\frac{k_i^z}{k_p^z} \rightarrow \infty$, one has that $\left[\frac{(k_i^z)}{(k_p^z)}\right]  \eta_z \rightarrow  v_z$ and (\ref{eq:etadynamicsz})--(\ref{eq:vdynamicsz}) reduces to the first-order system
\begin{eqnarray}
\dot{v}_z &=& -k_p^z  v_z + k_p^z (u_z+d_z)\,.
\end{eqnarray}
\endproof
\end{rmk}


\subsection{The Application Fits the Class of Plants (\ref{eq:planteta})--(\ref{eq:plantoutput})}\label{sec:UAVsimple}

Via experiments and simulation, we have verified that a first-order linear system can capture the main UAV's dynamics, for low velocity, while a relative degree one linear system, can capture the main behavior for medium velocities. For higher velocities, this model reduction fails. The parameters are given in Section~\ref{sec:simu}. 

\begin{figure}[h!]
\begin{center}
\includegraphics[width=\linewidth]{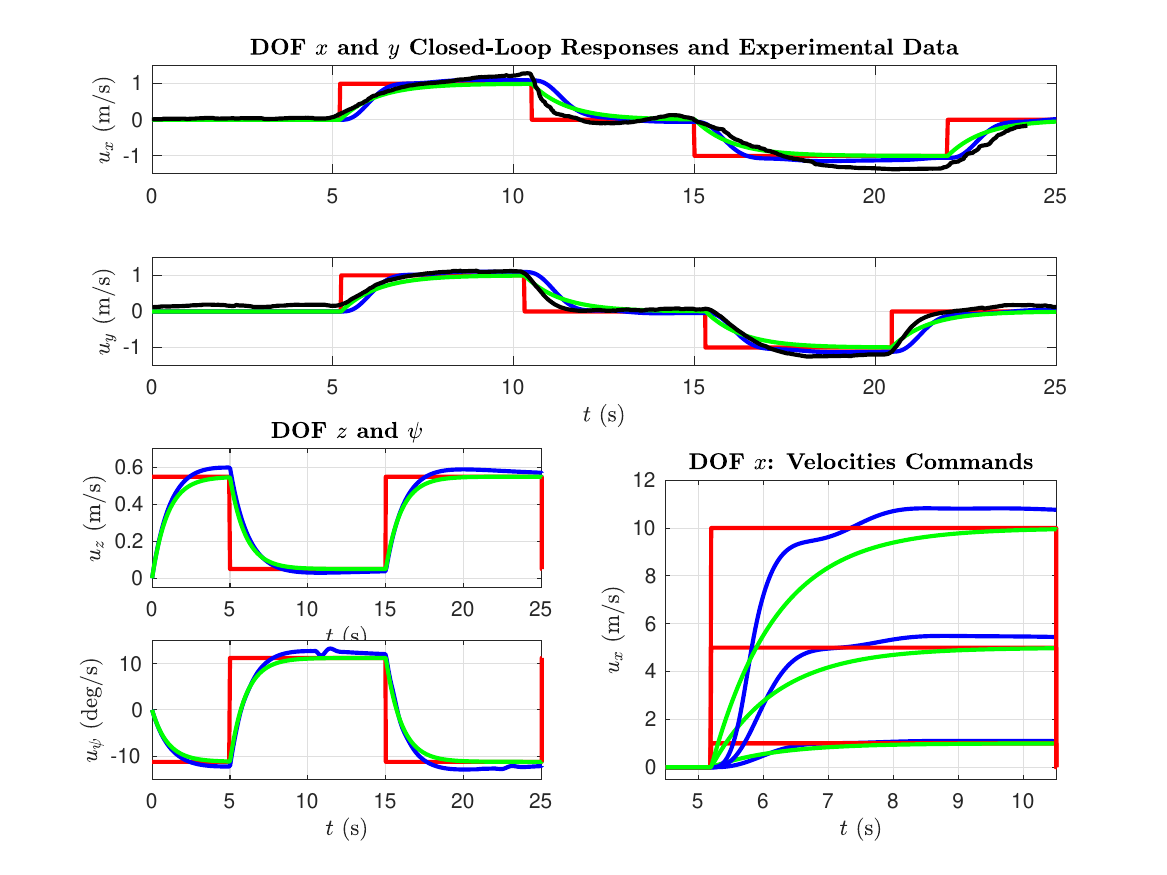}
\caption{Simulations based on the full UAV dynamics model in comparison with the simplified first-order model and validation of the simulator via experiments with the DJI M600.} 
\label{fig:comparison1stmodel}		
\end{center}
\end{figure}

The step responses are illustrated in Figure~\ref{fig:comparison1stmodel}, where we can see a reasonable match with a simple first-order system with a transfer function 
$G(s)=\frac{1}{s+1}$,
for all degrees of freedom, with some coupling disturbances among the subsystems. Note that, with the full dynamic model, this preliminary inner control is not so effective for decoupling the subsystems, in particular for the yaw subsystem illustrated at the bottom-left of Figure~\ref{fig:comparison1stmodel}, where one can observe a transient at $t=10$s due to the other channels step changes. As mentioned before, we also illustrated at the bottom-right of Figure~\ref{fig:comparison1stmodel} that a first-order model is not enough to represent the subsystem when the velocity increases. Experiments with the DJI M600 validate the simulations conducted. At the top of Figure~\ref{fig:comparison1stmodel}, one can verify the consistency of the simulator results (blue line) and the experimental results (black line), for the $x$ and $y$ subsystems. Due to the limited memory available during the experiment, only $x$ and $y$ subsystems were collected. Different experiments were conducted for the altitude and yaw subsystems, also leading to a consistent comparison with the simulation (curves not shown to save space).  
%
%
%
%

We restrict ourselves to the case of relative degree one (with order one {\bf or greater}) which is the simplest case amenable by pure Lyapunov design. 
Moreover, for medium/higher velocities and depending on the inner controllers, higher relative degree systems should be considered for representing the dynamic behavior from the velocity command input to the actual UAV velocity. Fortunately, our scheme can also deal with arbitrary relative degree plants, by using linear lead filters to estimate output time derivatives.

\section{Numerical Simulations and Experimental Results}\label{sec:simuexp}

In what follows, we presented the simulation results with the UAV's dynamic model, including the aerodynamic effects and the inner control loops, and the experimental evaluation with the DJI  M600 Pro hexacopter.  

\subsection{Numerical Simulations with the Full UAV Dynamic  Model}\label{sec:simu}

The aerodynamics parameters,  extracted from the literature \cite{CBT:2022},  are as follows: the thrust aerodynamic coefficient $k_{T_i}=0.0024$, in $N s^2/rad$, the  aerodynamic torque coefficient $c_{\tau}=0.57$, in $m rad/s^2$, the matrix coefficient  $K_{F_d}=\mbox{diag}\left(\left[\begin{array}{ccc}0.03 & 0.03 & 0.015\end{array}\right]\right)$ of the drag force on the structure, in $N  s^2/m^2$, and the matrix coefficient  $K_{F_{di}}=\mbox{diag}\left(\left[\begin{array}{ccc}1 & 1 & 1\end{array}\right]\right) (8 \times 10^{-6})$ of the propeller drag force, 
in $N s^2/(m rad)$.

To simplify the control allocation, without lost generality, we consider a quadrotor with the same weight, size, and geometry as the DJI M600. The UAV's parameters can be summarized as follows: the number of rotors $n_r=4$, the directions of rotation $s_1=1$, $s_2=-1$, $s_3=1$ and $s_4=-1$, the propeller half length $r=0.1$m (radius), the rotor displacement measured from the center of mass and along the horizontal plane $d=0.57$m, the UAV's inertia tensor (in $kg m^2$) $I_b=\mbox{diag}\left(\left[\begin{array}{ccc}0.4 & 0.4 & 0.74\end{array}\right]\right)$, the UAV's mass $m=10.5$kg, the propeller hub mass $m_i=0.1$kg ($i=1,2,3,4$), the propeller hub inertia tensor (in $kg m^2$) $I_i=\mbox{diag}\left(\left[\begin{array}{ccc}0.01 & 0.01 & 0.5 \times 10^{-5}\end{array}\right]\right)$. The arm length is, thus, $L=\sqrt{d^2+h^2}=0.57$m.
The inner control loops are based on state feedback linearization-based controllers with feedforward and integral actions, with control gains: $k_p^z=0$, $k_d^z=1$, $k_p^\psi=0.2$, $k_d^\psi=1$, $k_p^\phi=60$, $k_d^\phi=15$, $k_p^\theta=60$, $k_d^\theta=15$, $k_p^x=0$,  $k_d^x=1$, $k_p^y=0$ and $k_d^y=1$.

\begin{example}{\bf (The DSSC  Applied to Both UAV's Full and Simplified  Models)}\label{ex:simulDSSC}
For this example, a constant wind velocity $v_w=\left[\begin{array}{ccc} 8 & -8 & 8 \end{array}\right]^T$, in $m/s$, was added after $t=20s$. The effect can be observed only for the full UAV's model which incorporates the aerodynamic drag terms (blue lines).

All initial conditions were set at zero except the drone position $p_x(0)=10$m, $p_y(0)=10$m and $p_z(0)=10$m, and yaw angle $\psi(0)=(\pi/4)$rad ($45$deg).
The desired trajectories are: $p_{x_d}(t)=20 \sin(2 \pi/40 t)$, $p_{y_d}(t)=20 \cos(2 \pi/40 t)$, $p_{z_d}(t)=3 \sin(2 \pi/60 t)+5$, and $\psi_d(t)=-(\pi/4) \sin(2 \pi/40 t)+\pi/4$.
A first-order simplified UAV model (\ref{eq:plantx1})--(\ref{eq:plantoutput}), is considered perfectly known, for simplicity, with constants $a_p=a_p^n=1$ and $k_p=k_p^n=1$, for the four subsystems (the inverse dynamics is neglected). In this case, the nominal control can be chosen as in (\ref{eq:defunom}), with constants $c_e=(a_p^n-l_0) l_0/k_p^n$, $c_{m1}=-a_p^n$, $c_{m2}=-1$ and a time-varying coefficient $c_\sigma(t)=(l_0-a_p^n+1/\tau_m(t))/k_p^n$, which satisfies $|c_\sigma(t)| \leq (|l_0-a_p^n|+1/\delta_m)/k_p^n$.

Moreover, constant input disturbances were also added in each subsystem, after $t=20s$: $d=-0.8$, for the $x$-subsystem; $d=0.8$, for the $y$-subsystem; $d=0.2$, for the $z$-subsystem; and $d=0.1$, for the $\psi$-subsystem.

For all subsystems, the DSSC algorithm is implemented with $\tau_{av}$ constant and with the dynamic functions
$$k_o(t):=\kappa_{o} (|\sigma(t)|^{1/2} + \delta)\,, \quad \tau_m(t):=\kappa_m (|\sigma(t)|^{1/2} + \delta)\,,$$
where $\delta=1$ and $\kappa_m=4.0166$. Moreover, for the $x$ and $y$ subsystems, were selected the parameters $\kappa_{o}=110.651$, $\tau_{av}=0.03$ and $\varrho=1.5$. For the $z$-subsystem, were selected $\kappa_{o}=55.3255$, $\tau_{av}=0.06$ and $\varrho=0.5$. For the $\psi$-subsystem, were selected $\kappa_{o}=55.3255$, $\tau_{av}=0.06$ and $\varrho=0.15$.  The other DSSC´s parameter is $l_0=0.2$, for all subsystems. 
The closed-loop tracking performance of the DSSC is very similar to the performance of the original SSC, with the advantage that the steady-state values for $k_o(t)$ and $\tau_m(t)$ are obtained via ''online learning''. 
Figure~\ref{fig:DSSCwithFullUAVand2nd-kotaum} illustrates the time-varying behavior of the dynamic functions $\tau_{av}(t)$ and $\tau_m(t)$, where both increase when the disturbance acts after $t=20$s.
\begin{figure}[!ht]
\begin{center}
\includegraphics[width=\columnwidth]{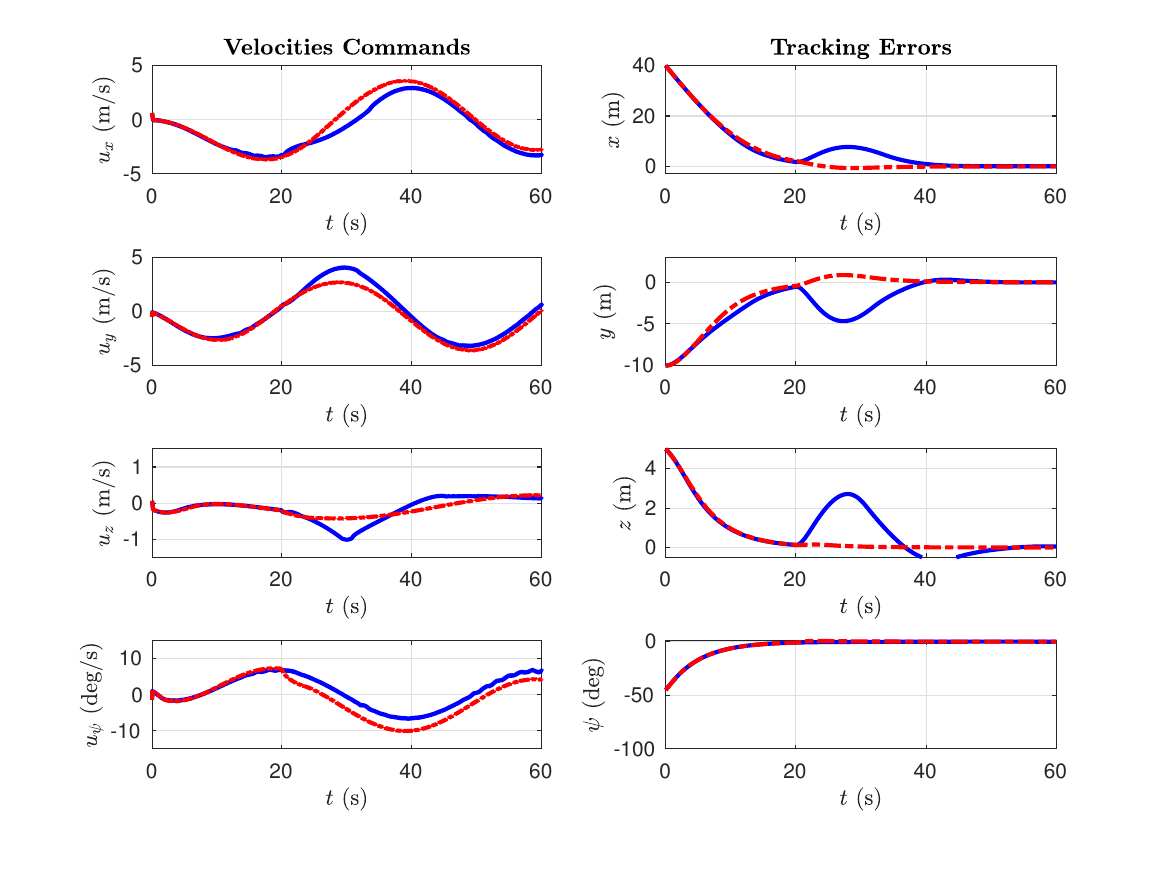}
\caption{Simulations of the DSSC with the full UAV dynamic model (blue line) and with the simplified model (red line). The  control efforts are in the left column, while the tracking errors are given in the right column.}
\label{fig:DSSCwithFullUAVand2nd-ue}		
\end{center}
\end{figure}
\begin{figure}[!ht]
\begin{center}
\includegraphics[width=\columnwidth]{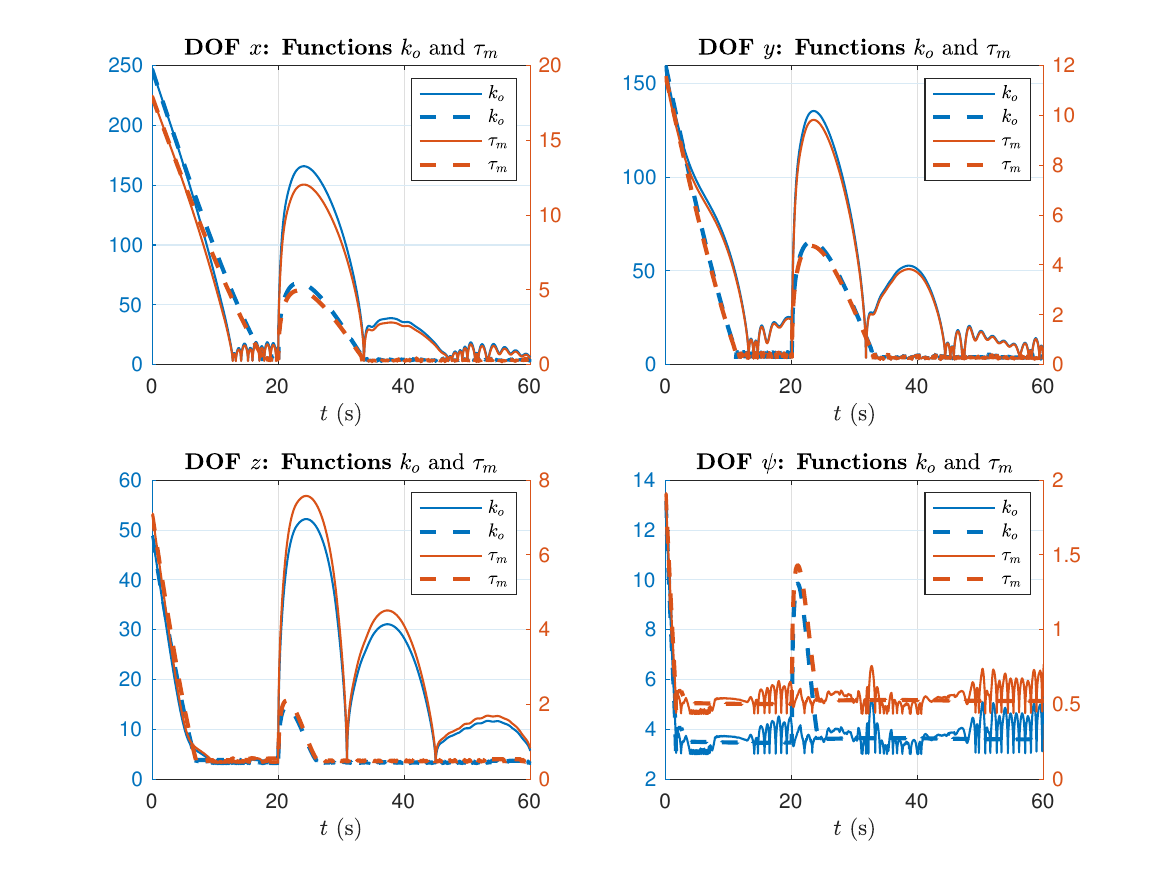}
\caption{Simulations of the DSSC with the full UAV dynamics model (blue line) and with the simplified  model (red line). The time-varying history of $k_o(t)$ and $\tau_m(t)$ are illustrated for the $4$ subsystems.}
\label{fig:DSSCwithFullUAVand2nd-kotaum}	
\end{center}
\end{figure}
In the left column of Figure~\ref{fig:DSSCwithFullUAVand2nd-ue}, one can see the velocity command reactions to compensate for the disturbances, after $t=20$s. Recall that the disturbances are different for the full UAV dynamic (wind disturbance) and for the simplified UAV model dynamics ($d$). However, before the disturbances ($t<20$), both tracking errors' behavior (right column) a very similar, except for a residual oscillation in the DSSC control signal (left column), when applied to the full UAV model case (blue line),  due to the  inner control loops (unmodelled dynamics-like effect). \endproof
\end{example}

\subsection{Experimental Results with the DSSC and the Standard STA}
\label{sec:exp}

\begin{figure}[h!]
\begin{center}
\includegraphics[width=.85\linewidth]{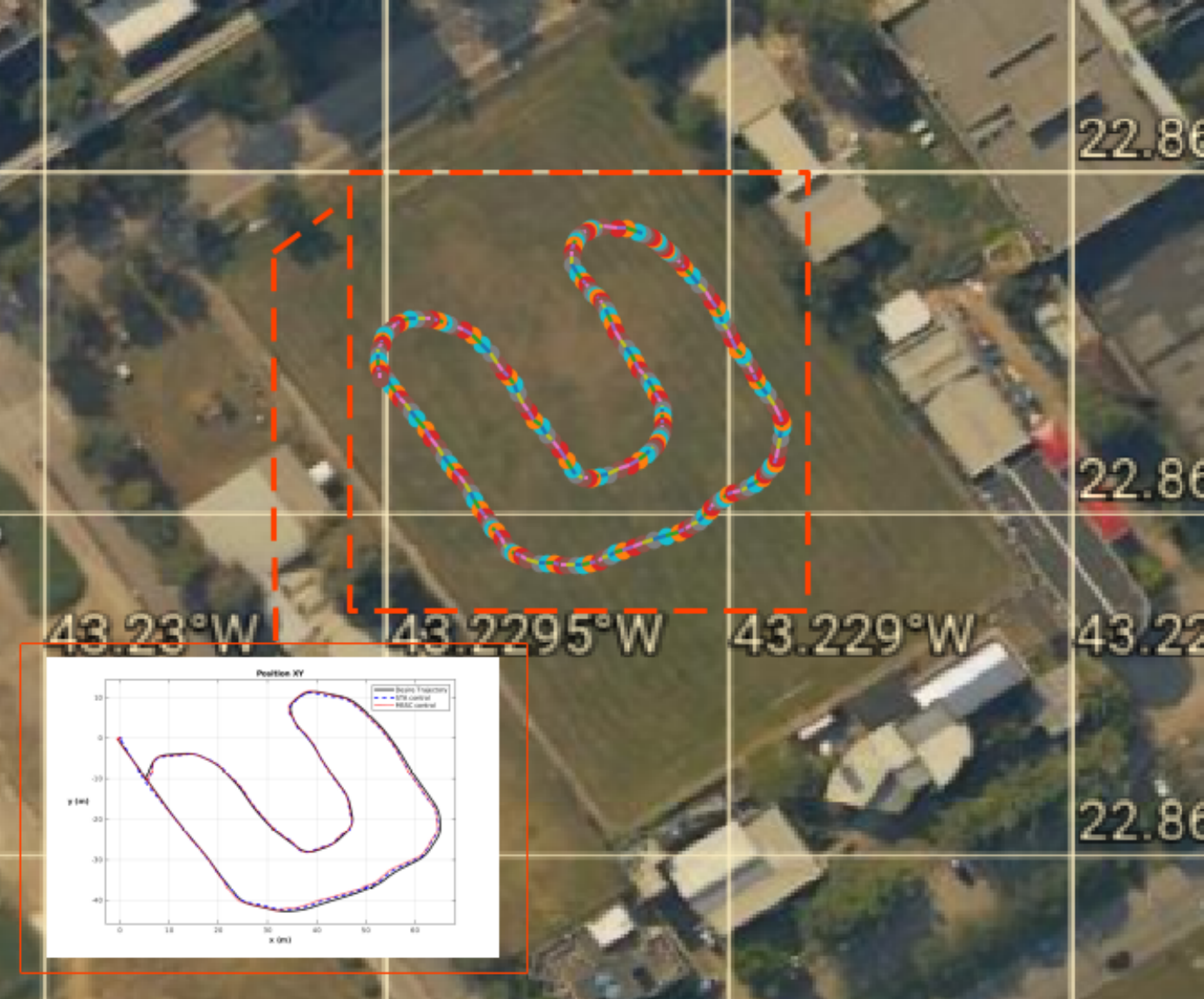}
\caption{Desired trajectory obtained via the developed Path Sketch Interface (PSI).}
\label{fig:trajpaper}		
\end{center}
\end{figure}

The desired trajectory was created to be executed in the field next to the laboratory (a soccer field), which is free of obstacles and barriers, at the Federal University of Rio de Janeiro. The path was obtained by using the Path Sketch Interface (PSI), a python interface with a satellite image from the area of interest that allows the users to choose the desired points, see Figure~\ref{fig:trajpaper}.
Then, a Matlab script converts the georeferenced points to the east-north-up (ENU)  reference system and generates a smooth trajectory version using a differentiable parametric curves approach. Finally, the controllers are developed using the Robotic Operation System (ROS) and C$++$. The ROS control node runs on an onboard Raspberry Pi 4 and loads the trajectory information generated by the Matlab script to execute the mission.

Our main purpose here is to experimentally evaluate the DSSC scheme in a real environment with the presence of real wind disturbances while ratifying that its closed-loop behavior during sliding mode approaches the STA. 

It must be highlighted that the same code implemented for all control laws works for the real-time implementation embedded in the UAV computer, as well as, in the simulator developed based on the full UAV model (\ref{eq:fullUAVmodel}) and in the  DJI Assistant $2$ Simulator.
The controllers were tested with and without wind disturbance in the DJI Assistant $2$. For the test with disturbance, the wind was added along the three axes ($x$, $y$ and $z$),  approximately at time $t=30s$. For the $x$ and $y$ axes, the wind speed of $8~m/s$ was introduced in the positive direction of movement. For the $z$ axis, the wind speed of $2~m/s$ was considered in the up direction. The results were omitted to save space.

After the test in the simulator, the  DSSC  and the STA were tested in a representative environment within the Federal University of Rio de Janeiro, a soccer field (Figure \ref{fig:trajpaper}), 
%
on the same day (April 20, 2022) and with the same wind conditions, i.e., a moderate wind with speed ranging from $5~m/s$ to $8~m/s$, according to the anemometer installed in the field. 

It was assumed that the nominal values for the uncertain parameters are (for all channels): $a_p^n=k_p^n=2$. The same DSSC's control parameters, as well as, the STA's parameters are used in all subsystems ($x,y,x$, and $\psi$). It was verified that a constant modulation function ($\varrho(t)=4$) was enough to deal with the uncertainties and the relative degree one output variable $\sigma$, in (\ref{eq:defsigma}), was implemented with  $l_0=2$.

The STA control  was tuned to ensure an acceptable performance in the real scenario, resulting in $\kappa_2=0.035$ and $\kappa_1 = 0.075$. The DSSC was implemented with 
$$\tau_{av}(t):=\frac{2}{k_o \kappa_1} |\sigma(t)|^{1/2}+\delta\,, \quad \tau_m(t):=\frac{\kappa_1}{2\kappa_2} |\sigma(t)|^{1/2}+\delta \,,$$
$k_o=10$ and $\delta=0.1$, for all  subsystems. The control gains of the STA and the DSSC's parameters were increased  in the experiments in comparison to the gains used in the DJI Assistant $2$ simulator. 

\begin{figure}[h!]
\begin{center}
\includegraphics[width=1\linewidth]{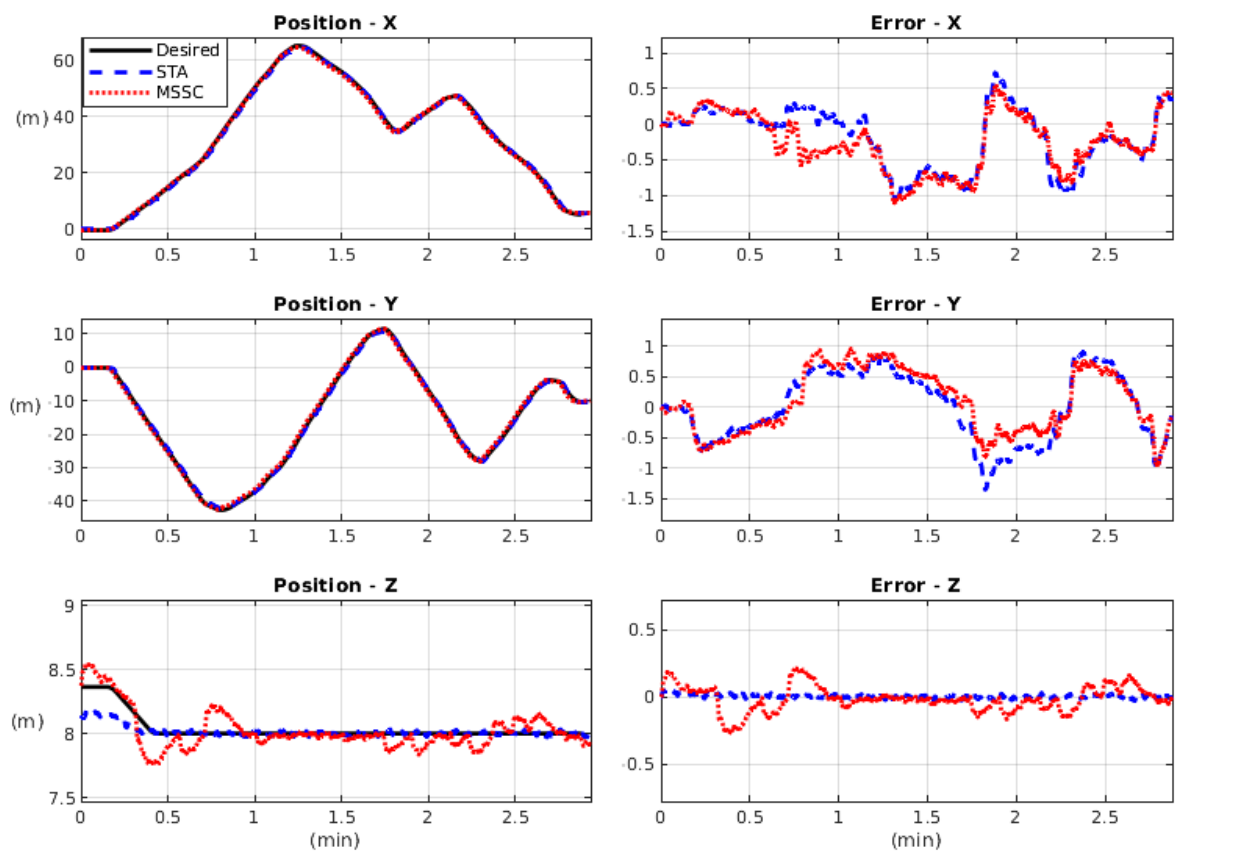}
\caption{Field Test. Trajectory tracking performance under STA  (dash blue) and DSSC (dot red) and the trajectory error along the three axes. The desired trajectory is illustrated in black.}
\label{fig:both_tracking}		
\end{center}
\end{figure}

Figure~\ref{fig:both_tracking} shows the closed loop tracking performance for both controls. In order to put in evidence the influence of the STA's gains, we have left the gain of the STA altitude control ($z$ axis) at the same level as in the simulation. This effect is clearly observed in the bottom of Figure~\ref{fig:both_tracking}, where the tracking error using the STA (red line) is significantly greater than the error using the DSSC scheme (blue line).  The left-bottom $ xy$ plot appearing in Figure~\ref{fig:trajpaper} illustrates the path tracking in the $xy$ axes. The corresponding control efforts are very similar (DSSC and STA), ratifying that its closed-loop behavior during sliding mode approaches the STA, but the curves are not shown to save space. 

\section{Conclusion}
\label{sec:conclusion}

The Dynamic Smooth Sliding Control (DSSC)  was proposed and successfully implemented on a real-scale UAV (hexacopter aircraft) for trajectory tracking in the presence of wind disturbances. Since the commercial hexacopter has  internal control loops not accessible by the user, a simulation was conducted based on a model for the UAV which includes the more relevant aerodynamic effects and the internal control loops. This allowed us to mimic the real UAV, as well as, the commercial simulator (which also has internal control loops not available for the designer), validating the developed simulator by experiments and simulations. 
It was verified that an approximation for the super twisting algorithm (STA) can be interpreted as the synthesized controller after the DSSC  achieves the sliding mode, for a particular choice of the dynamic functions employed in the smooth averaging filter and in the internal predictor. This approximation improves the robustness with respect to unmodelled dynamics due to a gain reduction near the origin.   
In addition, it was pointed out that both controllers had similar performances in the real experiment. The full closed-loop stability analysis was provided for the DSSC. Investigation of alternative dynamic functions for the  DSSC, closed-loop  stability analysis in the presence of parasitic or unmodeled dynamics, and a methodology to adapt the actual UAV simulation model for other types of UAVs are under development.

\bibliographystyle{elsarticle-num}
\bibliography{references}                

\appendix

\section{Proof of Theorem~\ref{theorem:DSSC}}
\label{ap:ProofTheorem}

The proof is carried out in two parts: before and after sliding mode takes place.

\medskip
{\bf Part A:  Analysis During the Reaching Phase}
\medskip

From the $\tilde{\sigma}$-dynamics (\ref{eq:sigmatildedynamicsDSSC}), one can obtain $\dot{\tilde{\sigma}} \tilde{\sigma} \leq-1/\tau_m \tilde{\sigma}^2   -\delta_\varrho |\tilde{\sigma}| \leq -\delta_\varrho |\tilde{\sigma}|$, 
since  the modulation function was designed in (\ref{eq:defvarrho}) to overcome the disturbance $d_0/k_o$, i.e., to satisfy $\varrho> |d_0|/k_o+\delta_\varrho/k_o$. The last inequality established the well know $\delta_\rho$-\textit{reachability condition} \cite{spurgeon}. Now, one can write
\begin{eqnarray}
 \dot{\tilde{\sigma}} \tilde{\sigma} &=& \frac{1}{2}\frac{d}{dt}\left[\tilde{\sigma}^2(t)\right]  \leq -\delta_\varrho |\tilde{\sigma}(t)|\,,\label{eq:sigmatildedynamicsDSSCproof5}
\end{eqnarray}
and integrating (\ref{eq:sigmatildedynamicsDSSCproof5}) from $t_0$ to $t \in [t_0,t_M)$, with $t \leq T_s$ and  $T_s:=t_0+|\tilde{\sigma}(t_0)|/\delta_\varrho$, it follows that 
\begin{eqnarray}
|\tilde{\sigma}(t)| &\leq& -\delta_\varrho (t-t_0)+|\tilde{\sigma}(t_0)| \leq |\tilde{\sigma}(t_0)| \,, \quad \forall t \in [t_0,t_M)\,, \quad \mbox{and} \quad t \leq T_s\,.\nonumber \label{eq:sigmatildedynamicsDSSCproof6}
\end{eqnarray}
It is clear that $\tilde{\sigma}(t_0)=0$ implies sliding mode at the manifold $\tilde{\sigma}(t)\equiv 0$, starting from the beginning, i.e., $\forall t \in [t_0,t_M)$, since $T_s=t_0+|\tilde{\sigma}(t_0)|/\delta_\varrho=t_0$ and the $\delta_\rho$-\textit{reachability condition} (\ref{eq:sigmatildedynamicsDSSCproof5}) is satisfied. In this case, finite-time escape cannot occur before sliding mode takes place. Thus, from now on, assume that $\tilde{\sigma}(t_0) \neq 0$.


Assuming that $t_M$ is finite, then there exists a finite $t^*$ ($t_0< t_M < t^*$) such that some close-loop signal escapes at $t=t^*$. Moreover, aiming to prove that finite-time escape cannot occur before sliding mode takes place, assume that $\tilde{\sigma}(t) \neq 0$, $\forall t \in [t_0,t^*]$.


%


Due to the unboundedness observability property of the closed-loop system, finite-time escape can occur if and only if the output $\sigma=\dot{e}+l_0 e$ escapes in finite-time. In addition, since the $\delta_\rho$-\textit{reachability condition} holds, then $\tilde{\sigma}$ is uniformly norm bound in the time interval $ [t_0,t^*]$.

Then, $\hat{\sigma}(t)=\sigma(t)-\tilde{\sigma}(t)$ must also escapes at $t=t^*$ and
$\lim_{t \rightarrow t^*} |\hat{\sigma}(t)| = \infty$.
However, at this point, $\hat{\sigma}$ can escape  to infinity while oscillating around zero (and switching sign), or monotonically with a fixed sign. The first case does not occur. Indeed, since $\varrho$ in (\ref{eq:defvarrho}) satisfies  $\varrho \geq |u_0^{av}|$, then the term $[-u_0^{av}+u_0]=[-u_0^{av}+\varrho \mbox{sgn}(\tilde{\sigma}(t_0))]$, appearing in the predictor  $\hat{\sigma}$-dynamics (\ref{eq:predictordynamicsDSSC}), has the same sign as $\tilde{\sigma}(t_0)$, where we have used the fact that $\mbox{sgn}(\tilde{\sigma}(t))=\mbox{sgn}(\tilde{\sigma}(t_0))$. Therefore, one can write the $\hat{\sigma}$-dynamics as
$$\tau_m(t) \dot{\hat{\sigma}}=- \hat{\sigma} + \tau_m(t) k_o(t) |-u_0^{av}+u_0|  \mbox{sgn}(\tilde{\sigma}(t_0))\,,$$
and $\hat{\sigma}$ cannot escape in finite-time oscillating and crossing zero, since the input  of the $\hat{\sigma}$-dynamics  has the fixed sign $\mbox{sgn}(\tilde{\sigma}(t_0))$ in the interval $[t_0,t^*]$.
So, there exists $T_0\in [t_0, t^*]$ such that $\hat{\sigma}(t) \neq 0$, $\forall t \in [T_0,t^*]$, escaping  monotonically with a fixed sign.

However, for $\mbox{sgn}(\tilde{\sigma}(t_0))=1$, one has $\tilde{\sigma}(t)>0$, $\sigma(t) > \hat{\sigma}(t)$, $\lim_{t \rightarrow t^*} \hat{\sigma}(t) = +\infty$ and the  strictly inequality
$q(t):=\frac{\sigma(t)}{\hat{\sigma}(t)}>1$
%
holds. Now, note that the quotient  
$q(t)=\frac{\sigma(t)}{\hat{\sigma}(t)}=1+\frac{\tilde{\sigma}(t)}{\hat{\sigma}(t)}$
and 
$\lim_{t \rightarrow t^*} \frac{\tilde{\sigma}(t)}{\hat{\sigma}(t)} = 0$,
%
%
%
where we have used the facts that $\tilde{\sigma}(t)$ is uniformly norm bounded in the closed time interval $[t_0,t^*]$ and $\lim_{t \rightarrow t^*} \hat{\sigma}(t) = +\infty$. Thus, one can further write
%
$\lim_{t \rightarrow t^*} q(t) = 1$,
which is a contradiction since $q(t)$ is strictly greater than one, $\forall t \in [t_0,t^*]$. 
For $\mbox{sgn}(\tilde{\sigma}(t_0))=-1$,  one has $\tilde{\sigma}(t)<0$, $k_\sigma\sigma(t) < \hat{\sigma}(t)$, $\lim_{t \rightarrow t^*} \hat{\sigma}(t) = -\infty$ and the inequality
$q(t):=\frac{k_\sigma\sigma(t)}{\hat{\sigma}(t)}<1$
holds in the closed time interval $[t_0,t^*]$. Analogously, the quotient $q(t)$ also satisfies $\lim_{t \rightarrow t^*} q(t) = 1$, 
which again is a contradiction since $q(t)$ is strictly less than one, $\forall t \in [t_0,t^*]$.

Finally, one can conclude that sliding mode occurs before any closed-loop signal escapes in finite time. However, finite-time escape is not precluded after sliding mode takes place. To complete the proof, we will evoke 
the Small Gain Theorem.

\medskip
{\bf Part B: Analysis in Sliding Mode}
\medskip

From {\bf Part (a)}, there exists a finite time $t_s \in [0,t_M)$ such that, $\forall t \in [t_s , t_M)$, sliding mode occurs, i.e.,  the sliding variable $\tilde{\sigma}(t)$ becomes identically null.

During sliding mode, the synthesized DSSC  law is given by $\bar{u}=\hat{u}_{vgsta}+C_s$, with $C_s:=\bar{u}(t_s)+\kappa_1(t_s) \hat{\phi}_1(t_s)$ and $\hat{u}_{vgsta}$ in (\ref{eq:uDSSCslidingNEW}), leading to
\begin{equation}
\bar{u}(t)= -\kappa_1(t) \hat{\phi}_1(t)-\int_{t_s}^t \kappa_2(\tau) \hat{\phi}_2(\tau) d\tau + C_s\,.\label{eq:uDSSCslidingNEWnew}
\end{equation}
Then, with $u=\bar{u}$, defining the auxiliary variable 
\begin{equation}
z:=-k_p\int_{t_s}^t \kappa_2(\tau) \hat{\phi}_2(\sigma(\tau)) d\tau +\sigma_a+k_p C_s\,,\label{eq:defz}
\end{equation}
with $\sigma_a$ in (\ref{eq:defsigmaa2}) and $\dot{\sigma}_a$ in (\ref{eq:defDotdsigmaa}),
%
%
the {\bf closed-loop system during sliding mode}  can be written as ($\forall t \in [t_s , t_M)$)
 \begin{eqnarray}
 \dot{e} &=& -l_0 e+ \sigma\,,\label{eq:planteAnalisys} \\
 \dot{\sigma}&=&-k_p \kappa_1 \hat{\phi}_1+\beta_1+z\,, \label{eq:plantsigmaAnalisys} \\
 \dot{z} &=& - k_p \kappa_2 \hat{\phi}_2 +\beta_2 +\beta_e + \beta_m\,, \label{eq:plantzAnalisys}
 \end{eqnarray}
%
with $\beta_1$ in (\ref{eq:defbeta1}), $\beta_2$ in (\ref{eq:defbeta2}), $\beta_e$ in (\ref{eq:defbetae}), $\beta_m$ in (\ref{eq:defbeta}),
where we have 
used the fact that $C_s$ is a constant.
As in \cite{morenoVGSTA1}, an additional transformation will be useful for the convergence analysis and gains design. Defining 
$$\zeta:=\left[\begin{array}{cc}
    \zeta_1 &  \zeta_2
\end{array}\right]=\left[\begin{array}{cc}
    \hat{\phi}_1 &  z
\end{array}\right]\,,$$
and noting that $\dot{\zeta}_1=\hat{\phi}_1^{'} \dot{\sigma}$ and $\hat{\phi}_2=\hat{\phi}_1^{'} \hat{\phi}_1$, we rewrite (\ref{eq:plantsigmaAnalisys}) and (\ref{eq:plantzAnalisys}) as $\dot{\zeta}_1=\hat{\phi}_1^{'}\left[-(k_p \kappa_1 - \alpha_1) \hat{\phi}_1 + z\right]$ and $\dot{\zeta}_2= \hat{\phi}_1^{'} \left[- (k_p {\kappa}_2 -{\alpha}_2) \hat{\phi}_1\right]+\beta_e + \beta_m$, respectively, where $\alpha_1$ and $\alpha_2$ are treated as  uncertain disturbances functions defined by
\begin{equation}
\alpha_1\hat{\phi}_1:=\beta_1 \quad \mbox{and} \quad \alpha_2\hat{\phi}_2:=\beta_2 \,, \quad \forall \sigma \neq 0\,,\label{eq:defalpha12}
\end{equation}
and $\alpha_1=\alpha_2=0$, for $\sigma=0$.
Finally, the closed-loop dynamics during sliding mode can be written in the compact form: 
%
%
%
\begin{eqnarray}
\dot{e} &=& -l_0 e+ \sigma\,,\label{eq:eproofA} \\
\dot{\zeta}&=& = \hat{\phi}_1^{'} A(\sigma,e,t) \zeta + B (\beta_e+\beta_m)\,,\label{eq:z2proofZetaA} 
\end{eqnarray}
where
$$A(\sigma,e,t):=\left[\begin{array}{cc}-(k_p \kappa_1 - \alpha_1)  & 1\\ - (k_p {\kappa}_2 -{\alpha}_2)  & 0\end{array}\right]\,, \quad B:=\left[\begin{array}{cc}0 & 1\end{array}\right]^T\,,$$
%
$\beta_e$ in (\ref{eq:defbetae}), $\beta_m$ in (\ref{eq:defbeta}) and $\alpha_1$ and $\alpha_2$ in (\ref{eq:defalpha12}).  
Similarly to \cite{morenoVGSTA1}, consider the Lyapunov function candidate
\begin{equation}
V(\zeta):=\zeta^T P \zeta\,, \quad P:=\left[\begin{array}{cc}\gamma k_p  & -2\epsilon \\ - 2\epsilon  & 1\end{array}\right]\,,
    \label{eq:defV}
\end{equation}
where $\gamma, \epsilon>0$ are design constants and $k_p$ is the plant's uncertain HFG (thus, $P$ is an uncertain matrix). Then, one can obtain 
\begin{equation}
\dot{V}=-\hat{\phi}_1^{'} \zeta^T Q \zeta+2 \zeta^{T} P B (\beta_e+\beta_m)\,,
    \label{eq:defdotV}
\end{equation}
where $Q:=-\left(A^T P +PA\right)$. 
%
The variable gains are designed (Table~\ref{tab:par}) in order to assure that matrix $Q-2 \epsilon I$ is positive definite, see \ref{sec:GainsLyap}. Now, with $Q-2 \epsilon I>0$ and reminding that $\hat{\phi}_1^{'} =  \phi_a \left[\frac{(|\sigma|^{1/2}+2\delta)}{2(|\sigma|^{1/2}+\delta)^2}\right]+\phi_b$, then one can write
\begin{equation}
   -\hat{\phi}_1^{'} \zeta^T Q \zeta \leq -2 \epsilon \frac{1}{\mu}\|\zeta\|^2 - 2 \epsilon \phi_b \|\zeta\|^2 \leq -2 \epsilon \frac{1}{\mu}\frac{V}{\lambda_{max}\{P\}}  - 2 \epsilon \phi_b \frac{V}{\lambda_{max}\{P\}} \,, \label{eq:dV1}
\end{equation}
%
where 
$$\mu:=\frac{2}{\phi_a} \left[\frac{(|\sigma|^{1/2}+\delta)^2}{(|\sigma|^{1/2}+2\delta)}\right]\,, \quad \phi_a\,, \delta>0\,,$$
%
and the Rayleigh quotient was applied.
%
To simplify the analysis at the cost of being more conservative and losing the capability of achieve the prescribed finite-time convergence for a residual set, we disregard this negative term $-2 \epsilon \frac{1}{\mu}\frac{V}{\lambda_{max}\{P\}}$ in (\ref{eq:dV1}), leading to 
\begin{equation}
   -\hat{\phi}_1^{'} \zeta^T Q \zeta \leq  - 2 \epsilon \phi_b \frac{V}{\lambda_{max}\{P\}} \,. \label{eq:dV1auxEXP}
\end{equation}
In addition, from (\ref{eq:boundonde}) and (\ref{eq:boundonbeta}), one can write $|\beta_e+\beta_m| \leq \|C_\eta A_\eta\| \|\eta\|+\kappa_e |e| + \bar{\beta}_m$, where $\bar{\beta}_m:=\bar{k}_p k_{d5}|y_m|+\bar{k}_p |\dot{u}_m^n| + |\ddot{\sigma}_m|+(l_0+\bar{a}_p) |\ddot{y}_m|+(\bar{k}_p k_{d4}+\|C_\eta B_\eta\|) |\dot{y}_m|+ \bar{k}_p (\alpha_{d2}+\alpha_{d3})$.
%
Then, the term $2 \zeta^{T} P B (\beta_e+\beta_m)$ in (\ref{eq:defdotV}) satisfies
\begin{equation}
|2 \zeta^{T} P B (\beta_e+\beta_m)| \leq \frac{2 \|PB\|}{\lambda_{min}^{1/2}\{P\}} V^{1/2} (\|C_\eta A_\eta\| \|\eta\|+\kappa_e |e| + \bar{\beta}_m)\,,\label{eq:boundinput}
\end{equation}
where  the relationship $\|\zeta\| \leq \frac{V^{1/2}}{\lambda_{min}^{1/2}\{P\}}$ was used. Hence, one can directly obtain the inequality $\dot{V} \leq -  \frac{2 \epsilon \phi_b}{\lambda_{max}\{P\}}V + \frac{2 \|PB\|}{\lambda_{min}^{1/2}\{P\}} V^{1/2} (\kappa_\eta |C_\eta \eta|+{\kappa}_e |e|+\bar{\beta}_m)$, or, equivalently, 
\begin{equation}
\dot{W}_v \leq -  \frac{\epsilon \phi_b}{\lambda_{max}\{P\}} W_v + \frac{\|PB\|}{\lambda_{min}^{1/2}\{P\}}  (\|C_\eta A_\eta\| \|\eta\|+{\kappa}_e |e|+\bar{\beta}_m)\,,\label{eq:dVineqEXP}
\end{equation}
%
where $W_v:=V^{1/2}$.

Moreover, reminding that $\dot{y}=[-l_0 e+\sigma+\dot{y}_m]$, then from (\ref{eq:planteta}) the inverse dynamics is given by $\dot{\eta}=A_\eta \eta + B_\eta [-l_0 e+\sigma+\dot{y}_m]$. Thus, one can put together the $e$-dynamics ($\dot{e} = -l_0 e+\sigma$) and the inverse dynamics and write 
\begin{eqnarray}
\dot{x}_\eta&=&A_{x\eta} x_\eta + B_\sigma \sigma + B_m  \dot{y}_m\,,\quad  x_\eta:=\left[\begin{array}{cc}\eta^T  & e\end{array}\right]^T\,,
\end{eqnarray}
where 
$A_{x\eta}:=\left[\begin{array}{cc}A_\eta  & -l_0 B_\eta\\0 & -l_0\end{array}\right]$, $B_\sigma:=\left[\begin{array}{cc}B_\eta^T  & 1\end{array}\right]^T$ and $B_m:=\left[\begin{array}{cc}B_\eta^T  & 0\end{array}\right]^T$.
Then, defining $V_\eta:=x_\eta^T P_\eta x_\eta$, with $P_\eta=P_\eta^T>0$ satisfying $A^T_{x \eta} P_\eta + P_\eta A_{x\eta}=-2I$, since $A_{x\eta}$ is a Hurwitz matrix, then one can obtain
$$\dot{V}_\eta \leq -\frac{2}{\lambda_{max}\{P_\eta\}} V_{\eta}+2\|P_\eta B_\sigma\|\frac{V_\eta^{1/2}}{\lambda_{min}^{1/2}\{P_\eta\}}|\sigma|+2\|P_\eta B_m\|\frac{V_\eta^{1/2}}{\lambda_{min}^{1/2}\{P_\eta\}}|\dot{y}_m|\,,$$
since $\|x_{x\eta}\| \leq \frac{V_\eta^{1/2}}{\lambda_{min}^{1/2}\{P_\eta\}}$. Equivalently, with $W_\eta:=V_\eta^{1/2}$, one has
\begin{eqnarray}
\dot{W}_\eta &\leq& -\frac{1}{\lambda_{max}\{P_\eta\}} W_{\eta}+\frac{\|P_\eta B_\sigma\|}{\lambda_{min}^{1/2}\{P_\eta\}}|\sigma|+\frac{\|P_\eta B_m\|}{\lambda_{min}^{1/2}\{P_\eta\}}|\dot{y}_m|\,.\label{eq:dVetaineqEXP}
\end{eqnarray}
Now, reminding that 
$$|\hat{\phi}_1| \leq \|\zeta\| \leq \frac{V^{1/2}}{\lambda_{min}^{1/2}\{P\}} = \frac{W_v}{\lambda_{min}^{1/2}\{P\}}\,, \quad \mbox{and} \quad |\hat{\phi}_1(\sigma)|:=\left[\frac{\phi_a}{(|\sigma|^{1/2} + \delta)} + \phi_b\right] |\sigma|\,,$$
then $|\hat{\phi}_1(\sigma)|\geq \phi_b |\sigma|$ and 
%
%
$|\sigma| \leq \frac{W_v}{\phi_b \lambda_{min}^{1/2}\{P\}}$. 


%
%
From (\ref{eq:dVineqEXP}) and (\ref{eq:dVetaineqEXP}), one has the following pair of inequalities:
\begin{eqnarray}
    \dot{W}_\eta &\leq& -\frac{1}{\lambda_{max}\{P_\eta\}} W_{\eta}+\frac{\|P_\eta B_\sigma\|}{\phi_b \lambda_{min}\{P_\eta\}}W_v +\frac{\|P_\eta B_m\|}{\lambda_{min}^{1/2}\{P_\eta\}}|\dot{y}_m|\,,\label{eq:dVineqEXPb} \\
    \dot{W}_v &\leq& -  \frac{\epsilon \phi_b}{\lambda_{max}\{P\}} W_v + \frac{\|PB\| {(\kappa}_e+\|C_\eta A_\eta\| )}{\lambda_{min}^{1/2}\{P\} \lambda_{min}^{1/2}\{P_\eta\}} W_\eta  +\frac{\|PB\|}{\lambda_{min}^{1/2}\{P\}}\bar{\beta}_m\,,\label{eq:dVetaineqEXPb}
\end{eqnarray}
where we use the fact that $|e|, \|\eta\| \leq \|x_\eta\| \leq  W_\eta/\lambda_{min}^{1/2}\{P_\eta\}$.
%
Now, let $\bar{W}_v$ and $\bar{W}_\eta$ be the solutions of the differential equations corresponding to the equalities in (\ref{eq:dVineqEXPb})--(\ref{eq:dVetaineqEXPb}),
with initial conditions $\bar{W}_v(t_s)=W_v(t_s)$ and $\bar{W}_\eta(t_s)=W_\eta(t_s)$. Thus, by using the Comparison Lemma \cite{K:02}, one has $W_v\leq \bar{W}_v$ and $W_\eta\leq \bar{W}_\eta$, $\forall t \in [t_s,t_M)$. 

Now, the proof follows by using the Small-Gain Theorem \cite{small-gain} applied to the pair of differential equations corresponding to the equalities in (\ref{eq:dVineqEXPb})--(\ref{eq:dVetaineqEXPb}). From \ref{sec:SmallGain}, for $\phi_b$ sufficiently large so that 
%
%
$$
\phi_b^2 \geq \frac{\|P_\eta B_\sigma\| \|PB\| {(\kappa}_e+\kappa_\eta \|C_\eta\|)\lambda_{max}\{P\}\lambda_{max}\{P_\eta\}}{4\epsilon \lambda_{min}^{1/2}\{P\} \lambda_{min}^{3/2}\{P_\eta\}}
$$
%
%
%
one can, subsequently, conclude that: $|z|$ converges exponentially to a residual set of order $\mathcal{O}(1/\phi_b)$, $|\sigma|$ and $|e|$ converges exponentially to a residual set of order $\mathcal{O}(1/\phi_b^2)$ and 
finite-time escape is avoided in all closed-loop signals.\endproof

\section{Auxiliary Signals and Norm Bounds}\label{ap:ErrorDynamicsDisturbancesBounds}


By considering the partitions (\ref{eq:defdpartition}) and (\ref{eq:defunomGERAL}) of the plant input disturbance  $d=d_1(y,\dot{y},t)+d_2(y,t)+d_3(t)$ and the nominal control $u^n=u_p^n(e)+u_d^n(\sigma)+u_i^n(t)+u_m^n(t)$,  one can decompose  the disturbance  $d_\sigma$ in (\ref{eq:sigmadynforsliding}) as $d_\sigma:=\beta_1+\sigma_a$, leading to $\dot{\sigma}=k_p u+\beta_1+\sigma_a$, where $\beta_1$ is the $\sigma$-independent signal 
\begin{equation}
\beta_1:=k_p u_d^n(\sigma) + (l_0-a_p) \sigma +k_p d_1(y,\dot{y},t) \,,\label{eq:defbeta1}
\end{equation}
and $\sigma_a$ is the \textit{auxiliary signal}
\begin{equation}
\sigma_a:=(l_0-a_p) (-l_0 e+\dot{y}_m)+ k_p (u_p^n + u_i^n +  u_m^n) - \dot{\sigma}_m+ k_p (d_2+d_3)-C_\eta \eta\,.\label{eq:defsigmaa2}
\end{equation}
%

%



%
The time derivative of the auxiliary signal $\sigma_a$ is $\sigma$-dependent, but can be decomposed in three signals: (i) $\beta_2$, which is $\sigma$-dependent; (ii) $\beta_e$, which is $e$-dependent; and $\beta_m$, which is an exogenous uniformly norm bounded time-varying signal.

In fact, $\dot{\sigma}_a$ can be written as
%
%
$\dot{\sigma}_a:=-(l_0-a_p)l_0 \dot{e} + k_p \frac{d u_p^n(e)}{d e} \dot{e} +k_p \bar{u}^n_i + k_p \dot{u}_m^n - \ddot{\sigma}_m+(l_0-a_p) \ddot{y}_m+ k_p \dot{d}_2+k_p \dot{d}_3-C_\eta \dot{\eta}$.
%
Now, since $\dot{d}_2=\frac{\partial d_2(y,t)}{\partial y} \dot{y}+\frac{\partial d_2(y,t)}{\partial t}$, $\dot{y}=-l_0 e+\sigma+\dot{y}_m$, 
%
%
$\dot{e}=-l_0 e + \sigma$ and $\dot{\eta}= A_\eta \eta + B_\eta \dot{y}$, one can write 
\begin{equation}
\dot{\sigma}_a:=\beta_2 +\beta_e + \beta_m\,,
    \label{eq:defDotdsigmaa}
\end{equation}
where
\begin{equation}
\beta_m:=k_p \dot{u}_m^n- \ddot{\sigma}_m+(l_0-a_p) \ddot{y}_m+k_p \frac{\partial d_2(y,t)}{\partial y} \dot{y}_m+k_p \dot{d}_3 -C_\eta B_\eta \dot{y}_m\,,\label{eq:defbeta}
\end{equation}
\begin{equation}
\beta_e:=-l_0 \left[-(l_0-a_p)l_0 +
 \frac{d u_p^n(e)}{d e}  -k_p l_0 \frac{\partial d_2(y,t)}{\partial y}-C_\eta B_\eta  \right] e- C_\eta A_\eta \eta + k_p \frac{\partial d_2(y,t)}{\partial t}\,,
\label{eq:defbetae}
\end{equation}%
\begin{equation}
\beta_2:=\left[-(l_0-a_p)l_0 +
k_p \frac{d u_p^n(e)}{d e} + k_p \frac{\partial d_2(y,t)}{\partial y} - C_\eta B_\eta \right] \sigma+k_p \bar{u}^n_i\,.
\label{eq:defbeta2}
\end{equation}%

\subsection{Auxiliary Norm Bounds}\label{ap:DisturbancesBounds}

In this section, the norm bounds for the $\beta$-terms are obtained. 
%
%
From Assumption {\bf (A4)}, one can obtain a non-negative constant $\bar{c}_\sigma$, such that $|k_p u_d^n + (l_0-a_p) \sigma| \leq \bar{c}_\sigma |\sigma|$. Hence, from Assumptions {\bf (A0)}--{\bf (A1)}, the disturbance $\beta_1$ in (\ref{eq:defbeta1}) satisfies $|\beta_1| \leq |k_p u_d^n + (l_0-a_p) \sigma| + |k_p d_1| \leq [\bar{c}_\sigma +\bar{k}_p (k_{d1} |y|+ k_{d2} |\dot{y}|+k_{d3})] |\sigma|$. Moreover, recalling that $y=e+y_m$ and $\dot{e}=-l_0 e + \sigma$, thus $\dot{y}=\dot{e}+\dot{y}_m=-l_0 e+\sigma+\dot{y}_m$ and 
%
\begin{equation}
|\beta_1| \leq (\bar{k}_{d1} |e|+ \bar{k}_{d2} |\sigma|+\bar{k}_{d3}) |\sigma|\,,
    \label{eq:boundondsigma1}
\end{equation}
with appropriate known constants $\bar{k}_{d1},\bar{k}_{d2},\bar{k}_{d3} \geq 0$, since $y_m$ and $\dot{y}_m$ are  uniformly norm-bounded signals.

From Assumption {\bf (A3)}, the term $k_p \dot{d}_3(t)$ of the signal $\beta_m$ defined in (\ref{eq:defbeta}) satisfies
$|k_p \dot{d}_3(t)| \leq \bar{k}_p \alpha_{d3}(t)$. Thus, $\beta_m$ is a uniformly norm-bounded signal which satisfies
\begin{equation}
|\beta_m| \leq \bar{k}_p |\dot{u}_m^n| + |\ddot{\sigma}_m|+(l_0+\bar{a}_p) |\ddot{y}_m|+\bar{k}_p k_{d4} |\dot{y}_m|+\bar{k}_p \alpha_{d3}+\|C_\eta B_\eta\| |\dot{y}_m|\,. \label{eq:boundonbeta}
\end{equation}
%
%
From Assumptions {\bf (A2)} and {\bf (A4)}, one can subsequently conclude that: (i) the term $\beta_e$ satisfies 
\begin{equation}
|\beta_e| \leq \|C_\eta A_\eta\| \|\eta\|+\kappa_e |e| + \bar{k}_p k_{d5}|y_m|+ \bar{k}_p \alpha_{d2}\,, \label{eq:boundonde}
\end{equation}
where $\kappa_e:=l_0 \left[(l_0+\bar{a}_p) l_0 +
 c_{e2}  +\bar{k}_p l_0 k_{d4} +\|C_\eta B_\eta\|\right]+\bar{k}_p k_{d5}$; and (ii) the term $\beta_2$ satisfies 
\begin{equation}%
|\beta_2| \leq ( k_\sigma+c_{i\sigma} |\sigma|+c_{ie}|e|)|\sigma|\,,
 \label{eq:boundondsigma}
\end{equation}
where $k_\sigma:=\left[(l_0+\bar{a}_p)l_0 +
\bar{k}_p c_{e2} + \bar{k}_p k_{d4} + \|C_\eta B_\eta\|\right]$.

\medskip
{\bf Norm Bounds $\rho_1$ and $\rho_2$}
\medskip


From (\ref{eq:defalpha12}) and  (\ref{eq:boundondsigma1}), one can write
$\phi_b |\sigma| |\alpha_1| \leq |\hat{\phi}_1| |\alpha_1|=|\beta_1| \leq  (\bar{k}_{d1} |e|+ \bar{k}_{d2} |\sigma|+\bar{k}_{d3}) |\sigma|$,
where the lower norm-bound comes from the fact that $\hat{\phi}_1 > \phi_b |\sigma|$, with $\hat{\phi}_1$ defined in (\ref{eq:defphi1}). Hence, one has $|\alpha_1| < \rho_1$, where 
$$\rho_1:=\frac{(\bar{k}_{d1} |e|+ \bar{k}_{d2} |\sigma|+\bar{k}_{d3})}{\phi_b}\,.$$
Analogously, from (\ref{eq:defalpha12}) and  (\ref{eq:boundondsigma}), one can write
$\phi_b^2 |\sigma| |\alpha_2|\leq |\hat{\phi}_1 \hat{\phi}_1^{'}| |\alpha_2|=|\beta_2| \leq  ( k_\sigma+c_{i\sigma} |\sigma|+c_{ie}|e|) |\sigma|$,
which leads to the norm bound $|\alpha_2| < \rho_2$, where 
$$\rho_2:= \frac{(k_\sigma + c_{i\sigma}|\sigma| + c_{ie}|e|)}{\phi_b^2}\,.$$



\section{Gain Functions Design}
\label{sec:GainsLyap}


The variable gains $\kappa_1$ and $\kappa_2$ are designed so that the matrix $Q$, appearing in (\ref{eq:defdotV}), satisfies $Q-2 \epsilon I>0$. One possibility is to set 
\begin{equation}
{\kappa}_2=2\epsilon \kappa_1+\gamma\,,
    \label{eq:k2design1}
\end{equation}
which leads to 
%
$$Q-2 \epsilon I=\left[\begin{array}{cc}2 (\gamma k_p-4\epsilon^2) k_p\kappa_1+4 \epsilon k_p \gamma-2 \gamma k_p \alpha_1 + 4 \epsilon {\alpha}_2 -2 \epsilon&  2 \epsilon \alpha_1 - {\alpha}_2\\
 2 \epsilon \alpha_1 - {\alpha}_2 & 2 \epsilon
\end{array}\right]\,,$$
that is positive definite for every value of $(t,e,\sigma)$ if 
%
%
%
\begin{equation}
    (\gamma k_p-4\epsilon^2) k_p\kappa_1>\frac{1}{4 \epsilon} (2 \epsilon \alpha_1 - {\alpha}_2)^2-2 \epsilon {\alpha}_2+ \gamma k_p(\alpha_1-2 \epsilon)+ \epsilon\,.\label{eq:k1designineq}
\end{equation}
%
%
It is clear that inequality (\ref{eq:k1designineq}) holds if the following one is valid
\begin{equation}
(\gamma k_p-4\epsilon^2) k_p\kappa_1 \geq \left[\frac{1}{4 \epsilon} (2 \epsilon \rho_1 +\rho_2)^2+2 \epsilon \rho_2+ \gamma k_p(\rho_1+2 \epsilon)+ \epsilon\right]\,,
    \label{eq:k1design}
\end{equation}
with $\gamma$ satisfying  $\gamma k_p-4\epsilon^2>0$,
$k_p$ being considered as an uncertain parameter and  $\rho_1$ and $\rho_2$ being known norm bounds for $\alpha_1$ and ${\alpha}_2$, respectively, obtained in what follows 
by using the available norm bounds for $\beta_1$ and $\beta_2$.
%
%
%

\medskip
{\bf Implementation of the  Variable Gain $\kappa_1$}
\medskip

The  gain $\kappa_1$ is designed to satisfy (\ref{eq:k1design}), which can be rewritten as
\begin{equation}
4 \epsilon k_p (\gamma k_p-4\epsilon^2)\kappa_1 >\left[ \rho^2+ 8 \epsilon^2 \rho_2+ 4 \epsilon \gamma k_p\rho_1 +  8 \epsilon^2 \gamma k_p + 4 \epsilon^2\right]\,,
    \label{eq:k1design1ineq}
\end{equation}
%
where 
$$\rho:=2 \epsilon \rho_1 +\rho_2 = \left[\frac{(\bar{k}_{d2} \phi_b + \bar{k}_p c_{i\sigma})}{\phi_b^2} |\sigma|+\frac{(\bar{k}_{d1} \phi_b + \bar{k}_p c_{ie})}{\phi_b^2} |e|+\frac{(\bar{k}_{d3} \phi_b + k_\sigma)}{\phi_b^2}\right]\,.$$ 
%
%
In addition, one sufficient condition to assure that (\ref{eq:k1design1ineq}) holds is given by
\begin{equation}
4 \epsilon k_p (\gamma k_p-4\epsilon^2)\kappa_1 >\left[ \rho^2+ (8 \epsilon^2+ 2 \gamma k_p)\rho +  8 \epsilon^2 \gamma k_p + 4 \epsilon^2\right]\,,
    \label{eq:k1design1ineqConservador}
\end{equation}
by noting that $\rho \geq \rho_2$ and $\rho \geq 2 \epsilon \rho_1$. Hence, 
since $\rho$, $\rho_1$ and $\rho_2$ are linearly related to $|\sigma|$ and $|e|$ and $8 \epsilon^2 \gamma k_p + 4 \epsilon^2$ is a constant, one can select $\kappa_1$ as
\begin{equation}
\kappa_1:= (\kappa_{a}|\sigma| + \kappa_{b}|e|+\kappa_{c})^2+\kappa_{d}:=\kappa^2+\kappa_d\,, \quad \kappa:=(\kappa_{a}|\sigma| + \kappa_{b}|e|+\kappa_{c})\,,
    \label{eq:k1design1}
\end{equation}
with positive constants $\kappa_{a},\kappa_{b},\kappa_{c}$ and $\kappa_{d}$ designed to assure that (\ref{eq:k1design1ineqConservador}) holds. One possible design is as follows. Firstly, restrict $\gamma$ (large enough) to satisfy
$4 \epsilon k_p (\gamma k_p-4\epsilon^2) >1$,
and restrict $\kappa_{c}$ to satisfy 
$\kappa_{c} > \frac{(8 \epsilon^2+ 2 \gamma k_p)}{[4 \epsilon k_p (\gamma k_p-4\epsilon^2)-1]}$.
Hence, one has that $\kappa
> \frac{(8 \epsilon^2+ 2 \gamma k_p)}{[4 \epsilon k_p (\gamma k_p-4\epsilon^2)-1]}$ and, consequently, 
%
%
%
%
%
$4 \epsilon k_p (\gamma k_p-4\epsilon^2) \kappa^2
> \kappa^2+ (8 \epsilon^2+ 2 \gamma k_p) \kappa$.
Secondly, restrict $\kappa_d$ to satisfies 
$$\kappa_d > \frac{(8 \epsilon^2 \gamma k_p + 4 \epsilon^2)}{4 \epsilon k_p (\gamma k_p-4\epsilon^2) }\,,$$
one has $4 \epsilon k_p (\gamma k_p-4\epsilon^2) (\kappa^2+\kappa_d)
> \kappa^2+ (8 \epsilon^2+ 2 \gamma k_p) \kappa+(8 \epsilon^2 \gamma k_p + 4 \epsilon^2)$. Thirdly, restricting $\kappa_{a}$, $\kappa_{b}$ and  $\kappa_{c}$ to satisfy
$$\kappa_{a}>\frac{(\bar{k}_{d2} \phi_b + \bar{k}_p c_{i\sigma})}{\phi_b^2}\,, \quad \kappa_{b}>\frac{(\bar{k}_{d1} \phi_b + \bar{k}_p c_{ie})}{\phi_b^2} \,, \quad \kappa_{c}>\frac{(\bar{k}_{d3} \phi_b + k_\sigma)}{\phi_b^2}\,,$$
one has $\kappa=(\kappa_{a}|\sigma| + \kappa_{b}|e|+\kappa_{c})
>\rho$, leading to the conclusion that  $\kappa^2+ (8 \epsilon^2+ 2 \gamma k_p) \kappa > \rho^2+ (8 \epsilon^2+ 2 \gamma k_p)  \rho$ and, consequently, 
$4 \epsilon k_p (\gamma k_p-4\epsilon^2) (\kappa^2+\kappa_d)
> \rho^2+ (8 \epsilon^2+ 2 \gamma k_p)  \rho+(8 \epsilon^2 \gamma k_p + 4 \epsilon^2)$.
Fourthly, restricting $\kappa_c$ to satisfy 
$$\kappa_c > \max\left\{\frac{(8 \epsilon^2+ 2 \gamma k_p)}{[4 \epsilon k_p (\gamma k_p-4\epsilon^2)-1]}\,, \frac{(\bar{k}_{d3} \phi_b + k_\sigma)}{\phi_b^2}\right\}\,,$$
one has that (\ref{eq:k1design1ineqConservador}) holds. 
%

\medskip
{\bf DSSC's Dynamic Function}
\medskip

Now, we will provide some additional restrictions to the parameters $\kappa_a$, $\kappa_b$  and $\phi_b$, so that the DSSC's dynamic functions $k_o \tau_{av}>0$ and $\tau_m>0$ are well-defined for all finite values of $\sigma, e$.
%
Two sufficient conditions for that are 
$$\left[\kappa_1^{'} \hat{\phi}_1 + \kappa_1 \hat{\phi}_1^{'}  \right]>0\,, \quad \mbox{and} \quad \left[\frac{\partial \kappa_1}{\partial e} [-l_0 e+\sigma]+\frac{\partial \kappa_1}{\partial t}+\kappa_2\hat{\phi}_1^{'}\right]>0\,.$$
From (\ref{eq:k1design1}), one has  $\kappa_1^{'}= \frac{\partial \kappa_1}{\partial \sigma} =2 (\kappa_{a}|\sigma| + \kappa_{b}|e|+\kappa_{c})\kappa_a \mbox{sgn}(\sigma)$, $\frac{\partial \kappa_1}{\partial e} = 2 (\kappa_{a}|\sigma| + \kappa_{b}|e|+\kappa_{c})\kappa_b\mbox{sgn}(e)$ and $ \frac{\partial \kappa_1}{\partial t}=0$. In addition, recall that $\hat{\phi}_1:=\frac{\phi_a \sigma}{(|\sigma|^{1/2} + \delta)} + \phi_b \sigma$ and  $\hat{\phi}_1^{'}= \phi_a \left[\frac{(|\sigma|^{1/2}+2\delta)}{2(|\sigma|^{1/2}+\delta)^2}\right]+\phi_b$.

Hence, one can directly conclude that $\kappa_1^{'} \hat{\phi}_1>0$, leading to $\left[\kappa_1^{'} \hat{\phi}_1 + \kappa_1 \hat{\phi}_1^{'}  \right]>0$. Moreover, since $\kappa_2=2\epsilon \kappa_1+\gamma$, one has
\begin{equation}
\left[\frac{\partial \kappa_1}{\partial e} [-l_0 e+\sigma]+\frac{\partial \kappa_1}{\partial t}+\kappa_2\hat{\phi}_1^{'}\right] = \left[\frac{\partial \kappa_1}{\partial e} [-l_0 e+\sigma]+(2\epsilon \kappa_1+\gamma)\hat{\phi}_1^{'}\right]>0\,, \nonumber
    \label{fgeq:k1design1}
\end{equation}
if the following inequality is valid
\begin{equation}
(2\epsilon \kappa_1+\gamma)\hat{\phi}_1^{'} > \left|\frac{\partial \kappa_1}{\partial e} [-l_0 e+\sigma]\right|\,.
    \label{fgeq:k1design2}
\end{equation}
On the other hand, 
$$\left|\frac{\partial \kappa_1}{\partial e} [-l_0 e+\sigma]\right| \leq 2 (\kappa_{a}|\sigma| + \kappa_{b}|e|+\kappa_{c})\kappa_b (l_0|e|+|\sigma|)\,,$$
and 
$$(2\epsilon \kappa_1+\gamma)\hat{\phi}_1^{'}=(2\epsilon [(\kappa_{a}|\sigma| + \kappa_{b}|e|+\kappa_{c})^2+\kappa_d]+\gamma)\hat{\phi}_1^{'}\,.$$
Now, if 
\begin{equation}
    (2\epsilon [(\kappa_{a}|\sigma| + \kappa_{b}|e|+\kappa_{c})^2+\kappa_d]+\gamma)\hat{\phi}_1^{'} > 2 (\kappa_{a}|\sigma| + \kappa_{b}|e|+\kappa_{c})\kappa_b (l_0|e|+|\sigma|)\,,\label{eq:ine}
\end{equation}
then the DSSC's dynamics functions are well-defined. Moreover, note that the following inequality is a sufficient condition to (\ref{eq:ine}) hold:
$$2\epsilon (\kappa_{a}|\sigma| + \kappa_{b}|e|+\kappa_{c})^2\hat{\phi}_1^{'} > 2 (\kappa_{a}|\sigma| + \kappa_{b}|e|+\kappa_{c})\kappa_b (l_0|e|+|\sigma|)\,,$$
or, the following one
$$\epsilon (\kappa_{a}|\sigma| + \kappa_{b}|e|+\kappa_{c}) \phi_b > \kappa_b (l_0|e|+|\sigma|)\,,$$
where we use the fact that 
$$\hat{\phi}_1^{'} =  \left[\phi_a \left[\frac{(|\sigma|^{1/2}+2\delta)}{2(|\sigma|^{1/2}+\delta)^2}\right]+\phi_b\right] > \phi_b\,.$$
Finally, one can guarantee that  the DSSC's dynamics functions are well-defined, by selecting  $\phi_b > \frac{l_0}{\epsilon}$ and $\kappa_a >\frac{\kappa_b}{l_0}$. This is summarized in  Table~\ref{tab:par}.

\section{Stability Analysis via Small-Gain Theorem}\label{sec:SmallGain}

The equalities in (\ref{eq:dVineqEXPb})--(\ref{eq:dVetaineqEXPb}) are given by
\begin{eqnarray}
    \dot{\bar{W}}_\eta &=& -\frac{1}{\lambda_{max}\{P_\eta\}} \bar{W}_{\eta}+\frac{\|P_\eta B_\sigma\|}{\phi_b \lambda_{min}\{P_\eta\}} W_v +\frac{\|P_\eta B_m\|}{\lambda_{min}^{1/2}\{P_\eta\}}|\dot{y}_m|\,,\label{eq:dVineqEXPbequality} \\
    \dot{\bar{W}}_v &=& -  \frac{\epsilon \phi_b}{\lambda_{max}\{P\}} \bar{W}_v + \frac{\|PB\| {(\kappa}_e+\kappa_\eta \|C_\eta\|)}{\lambda_{min}^{1/2}\{P\} \lambda_{min}^{1/2}\{P_\eta\}} W_\eta  +\frac{\|PB\|}{\lambda_{min}^{1/2}\{P\}}\bar{\beta}_m\,,\label{eq:dVetaineqEXPbequality}
\end{eqnarray}
which can be rewritten as
%
%
\begin{eqnarray}
    \dot{\mathcal{X}}_1 &=&  -\lambda_1 \mathcal{X}_1 +  \frac{g_1}{\phi_b}\mathcal{X}_2+ \mathcal{U}_1\,,\label{eq:deineqEXPbODE1SmallGain} \\
    \dot{\mathcal{X}}_2 &=& -  \lambda_2 \phi_b \mathcal{X}_2 + g_2 \mathcal{X}_1+ \mathcal{U}_2\,,\label{eq:dVineqFiniteTimebODE1SmallGain}
\end{eqnarray}
where $\mathcal{X}_1:=\bar{W}_\eta$, $\lambda_1:=\frac{1}{\lambda_{max}\{P_\eta\}}$, $\mathcal{U}_1:=g_1 (W_v-\mathcal{X}_2)/\phi_b+\bar{g}_1 |\dot{y}_m|$, $g_1=\frac{\|P_\eta B_\sigma\|}{\lambda_{min}\{P_\eta\}}$, $\bar{g}_1=\frac{\|P_\eta B_m\|}{\lambda_{min}^{1/2}\{P_\eta\}}$,
$\mathcal{X}_2:=\bar{W}_v$, $\lambda_2:=\frac{\epsilon}{\lambda_{max}\{P\}}$, $\mathcal{U}_2:=g_2 (W_\eta-\mathcal{X}_1)+\bar{g}_2 \bar{\beta}_m$, $g_2:=\frac{\|PB\| {(\kappa}_e+\kappa_\eta \|C_\eta\|)}{\lambda_{min}^{1/2}\{P\} \lambda_{min}^{1/2}\{P_\eta\}}$ and $\bar{g}_2:=\frac{\|PB\|}{\lambda_{min}^{1/2}\{P\}}$.
Thus, one can write 
\begin{eqnarray}
    \mathcal{X}_1 & \leq &  e^{-\lambda_1 (t-t_s)} \mathcal{X}_1(t_s) +   \frac{\gamma_1}{\phi_b} \|\mathcal{X}_2\|_{\infty} + \frac{1}{\lambda_1} \|\mathcal{U}_1\|_{\infty}\,,\label{eq:deineqEXPbODE1SmallGainIneq1} \\
     \mathcal{X}_2 & \leq & e^{-\lambda_2 \phi_b (t-t_s)} \mathcal{X}_2(t_s)  +  \frac{\gamma_2}{\phi_b} \|\mathcal{X}_1\|_{\infty} +\frac{1}{\lambda_2 \phi_b} \|\mathcal{U}_2\|_{\infty}\,,\label{eq:dVineqFiniteTimebODE1SmallGainIneq1}
\end{eqnarray}
where  $\gamma_1:=\frac{g_1}{\lambda_1}$ and $\gamma_2:=\frac{g_2}{\lambda_2}$.

Now, one can apply \cite[Theorem 2.1]{small-gain} and conclude that the system (\ref{eq:deineqEXPbODE1SmallGain}) and (\ref{eq:dVineqFiniteTimebODE1SmallGain}) with inputs $\mathcal{U}_1$ and $\mathcal{U}_2$, outputs $\mathcal{Y}_1=\mathcal{X}_1$ and $\mathcal{Y}_2=\mathcal{X}_2$ and states  $\mathcal{X}_1$ and $\mathcal{X}_2$ is IOpS, satisfying
\begin{eqnarray}
\|\mathcal{X}(t)\| &\leq& \beta_{\mathcal{X}}(\|\mathcal{X}(t_s)\|,t-t_s) + \alpha_{\mathcal{X}}(\|\mathcal{U}(t)\|_{\infty})\,,\label{eq:ResidualX}
\end{eqnarray}
for some class-$\mathcal{KL}$ function $\beta_{\mathcal{X}}$ and some class-$\mathcal{K}$ function $\alpha_{\mathcal{X}}$, provided that there exist $k_l\geq 0$ and class-$\mathcal{K}_\infty$ functions $\mathcal{F}_1(s)+s$ and $\mathcal{F}_2(s)+s$ such that
$$\mathcal{F}_2(\gamma_2 (\mathcal{F}_1(\gamma_1 s)+\gamma_1 s))+\gamma_2 (\mathcal{F}_1(\gamma_1 s)+\gamma_1 s) \leq s\,, \quad \forall s\geq s_l\,.
$$
In particular, for $\mathcal{F}_1(s)=\mathcal{F}_2(s)=s$ and $k_l=0$, one has the following condition $4\gamma_1 \gamma_2 \leq \phi_b^2$. 

Now, since $W_v < \bar{W}_v=\mathcal{X}_2$, one has
$[W_v(t)-\mathcal{X}_2(t)]<0$, $\forall t \in[t_s,t_M)$, 
from which one can, subsequently, conclude that $\mathcal{U}_1(t):=g_1 [W_v(t)-\mathcal{X}_2(t)]+\bar{g}_1 |\dot{y}_m(t)| < \bar{g}_1 |\dot{y}_m(t)|$, $\forall t \in[t_s,t_M)$ and  
$$\|\mathcal{U}_1\|_{\infty} < \bar{g}_1 \|\dot{y}_m\|_{\infty}\,.$$
Analogously, one has 
$$\|\mathcal{U}_2\|_{\infty} < \bar{g}_2 \|\bar{\beta}_m\|_{\infty}\,,$$
since $W_\eta < \bar{W}_\eta=\mathcal{X}_1$. Thus, $\|\mathcal{U}\|_{\infty} < \bar{g}_1 \|\dot{y}_m\|_{\infty}+\bar{g}_2 \|\bar{\beta}_m\|_{\infty}$ is uniformly bounded by a $\phi_b$-independent constant. Hence, from (\ref{eq:ResidualX}), one has that $\mathcal{X}(t), \mathcal{X}_1(t), \mathcal{X}_2(t)$ converge to a residual set independent of $\phi_b$ and the initial conditions. Now, backing to (\ref{eq:dVineqFiniteTimebODE1SmallGainIneq1}), one can conclude that $\mathcal{X}_2(t)$ converges to a residual set of order $\mathcal{O}(1/\phi_b)$.
Finally,  reminding that 
$$|\hat{\phi}_1|, |z| \leq \|\zeta\| \leq \frac{V^{1/2}}{\lambda_{min}^{1/2}\{P\}} = \frac{W_v}{\lambda_{min}^{1/2}\{P\}} \leq \frac{\mathcal{X}_2}{\lambda_{min}^{1/2}\{P\}}\,,$$
and $|\hat{\phi}_1|\geq \phi_b |\sigma|$, 
the one has that $|z|$ converges to a residual set of order $\mathcal{O}(1/\phi_b)$ and  $|\sigma|$ (and $|e|$) converges to a residual set of order $\mathcal{O}(1/\phi_b^2)$.\endproof

\end{document}